\documentclass[twocolumn,amsmath,amssymb,floatfix,superscriptaddress,aps,prl,10pt]{revtex4-2}
\usepackage[english]{babel}
\usepackage{fullpage}
\usepackage{amssymb}
\usepackage{graphicx,color}
\usepackage{bbm}
\usepackage{multirow}
\usepackage{bm}
\usepackage{mathrsfs}
\usepackage{commath}
\usepackage{stmaryrd}
\usepackage{color}
\usepackage{subfigure}
\usepackage{comment}
\usepackage[utf8]{inputenc}
\usepackage[normalem]{ulem}
\usepackage{amsmath}
\usepackage{graphicx}
\usepackage{float}
\usepackage{verbatim}
\usepackage{esint}
\usepackage{epstopdf}
\usepackage{braket}
\usepackage[ruled,lined]{algorithm2e}

\usepackage{hyperref}
\hypersetup{
  colorlinks   = true,    
  urlcolor     = blue,    
  linkcolor    = blue,    
  citecolor    = blue     
}

\makeatletter
\@ifundefined{textcolor}{}
{%
 \definecolor{BLACK}{gray}{0}
 \definecolor{WHITe}{gray}{1}
 \definecolor{ReD}{rgb}{1,0,0}
 \definecolor{GReeN}{rgb}{0,1,0}
 \definecolor{BLUe}{rgb}{0,0,1}
 \definecolor{CYAN}{cmyk}{1,0,0,0}
 \definecolor{MAGeNTA}{cmyk}{0,1,0,0}
  \definecolor{PURPLE}{cmyk}{0.4,0.8,0,0}
 \definecolor{YeLLOW}{cmyk}{0,0,1,0}
}

\def\d{{\rm d}}
\def\R{{\rm R}}

\def\hn{\hat{n}}

\def\hphi{\hat{\phi}}
\def\hpsi{\hat{\psi}}

\usepackage{epsfig}
\usepackage{dcolumn}
\usepackage{bm}

\makeatother

\begin{document}

\title{Single-worldline theory for a dissipative Mott transition}

\author{Oscar Bouverot-Dupuis}
\affiliation{Universit\'{e} Paris Saclay, CNRS, LPTMS, 91405, Orsay, France}
\affiliation{IPhT, CNRS, CEA, Universit\'{e} Paris Saclay, 91191 Gif-sur-Yvette, France}

\author{Alberto Rosso}
\affiliation{Universit\'{e} Paris Saclay, CNRS, LPTMS, 91405, Orsay, France}

\author{Laura Foini}
\affiliation{IPhT, CNRS, CEA, Universit\'{e} Paris Saclay, 91191 Gif-sur-Yvette, France}

\begin{abstract}
In one dimension at zero temperature, local baths with spectral exponent $s$ can stabilize a compressible, non-superfluid dissipative phase between the Mott insulator and Luttinger liquid. The dissipative-to-Mott transition is accessible neither to perturbative renormalization-group methods nor to the free-fermion description of the conventional Mott transition. Here, we show it is governed by the worldline of a single doped excitation, whose geometrical roughness determines the critical exponents. Without dissipation, this worldline undergoes Brownian motion, recovering $\beta=\nu=1/z=1/2$. Dissipation turns it into a long-range interacting interface, yielding continuously varying exponents $\beta=\nu=1/z=s-1$ for $1<s<3/2$, and $\beta=\nu=1/z=0$ for $s<1$. The same theory identifies $s=3/2$ as the threshold above which the dissipative phase disappears. Large-scale Monte Carlo simulations of both the single-worldline theory and the original many-body model quantitatively support these predictions.
\end{abstract}

\date{\today}

\maketitle

\paragraph{Introduction ---}

Coupling a quantum system to an environment is usually associated with decoherence, but it can also stabilize exotic phases of matter. This interplay is particularly subtle in one dimension, where strong quantum fluctuations make collective ordering and critical phenomena especially sensitive to additional long-range interactions \cite{Mermin_Wagner_1966}.

While dissipation-induced localization is well understood for a single quantum degree of freedom \cite{Leggett_1987,Schon_1990,Schmid_1983,Fisher_1985}, its interplay with many-body quantum criticality remains largely unexplored. The one-dimensional Mott transition, in which a conducting Luttinger liquid (LL) turns into a Mott insulator (MI) under the effect of a periodic potential \cite{Mott_1968,Fisher_1989_SI,Giamarchi_1997_Mott1D,Mott_2004}, provides a particularly sharp setting for this question. A natural realization of dissipation in this system is obtained by coupling each lattice site to an independent bath of harmonic oscillators with spectral exponent $s$ \cite{Cazalilla_2026,Ribeiro2024,Radzihovsky_2024,Malatsetxebarria_2013,Cai2014,Majumdar2023a,Bouverot_Dupuis_2024}, as in the Caldeira--Leggett framework \cite{Caldeira_1983,Leggett_1987}.

Dissipation qualitatively changes the Mott transition: away from commensurability it stabilizes a compressible but non-superfluid dissipative phase (DP) \footnote{This dissipative phase is also called a dissipative insulator (DI) in \cite{Malatsetxebarria_2013}, a bath induced Bose liquid in \cite{Cai2014}, a dissipative phase in \cite{Majumdar2023a,Majumdar_2023b}, and a dissipative fixed point (DFP) in \cite{Ovni_2025}. The dual phase where the bosonized phase fields $\phi$ and $\theta$ are interchanged is called a superconducting phase in \cite{Lobos_2009}, an incoherent transverse quantum fluid (iQTF) in \cite{kuklov_2024a,Kuklov_2024b}, an incoherent superfluid (iSF) in \cite{Radzihovsky_2024}, a dissipative superfluid in \cite{Ribeiro2024} and a dissipative Bose-Einstein condensate (D-BEC) in \cite{Cazalilla_2026}.}, separated from the LL by a BKT transition \cite{Malatsetxebarria_2013,Cai2014,Majumdar2023a,Bouverot_Dupuis_2024,bouvdup_2025_bosonised1d,Ovni_2025}. How this phase gives way to the MI, however, is an open problem, and no critical theory for the DP--MI transition exists as it is inaccessible to both perturbative and free-fermion methods that describe the conventional Mott transition \cite{Japaridze_1978,Pokrovsky_1979,Schulz_1980_CI,Giamarchi_1991}.

Here, we derive such a theory. As the MI is approached, the typical distance between doped excitations diverges, and we show that the critical properties of the DP--MI transition are encoded in the fluctuations of a single-excitation worldline. This maps the original many-body problem onto a classical one-dimensional interface with long-range interactions, from which the DP--MI critical exponents $\nu$, $\beta$ and $z$ are read off directly as geometrical quantities. For $s<3/2$, we find a new family of universality classes continuously indexed by $s$. Above $s=3/2$, the worldline behaves as a Brownian path and one recovers the exponents of the conventional LL--MI Mott transition. This establishes the topology of the full phase diagram, with the DP existing only for $s<3/2$.

\begin{figure}[h!]
    \centering
    \includegraphics[width=0.85\linewidth]{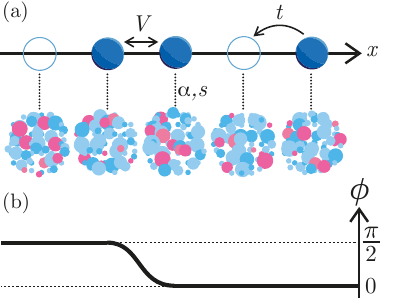}
    \caption{(a) Schematic representation of the dissipative system. A one-dimensional fermionic chain is coupled to independent local baths characterized by the spectral function $J(\omega) \simeq \alpha|\omega|^s$. (b) Through bosonization, a domain wall separating a region with one particle every even site from another with one particle every odd site is represented by a kink of amplitude $\pi/2$ in the field $\phi$.}
    \label{fig:model}
\end{figure}

\paragraph{Model ---}
We consider spinless fermions on a lattice (equivalently, an
XXZ spin chain) with each site coupled to an independent bath of harmonic oscillators $\hat{a}^{(\dagger)}_{j\gamma}$ (see Fig.~\ref{fig:model} (a)), described by the Hamiltonian
\begin{align}\label{eq:H}
    \hat{H}=\sum_j& \Big[ -t (\hpsi^\dagger_j \hpsi_{j+1} + \hpsi^\dagger_{j+1} \hpsi_j) + V \hn_j \hn_{j+1}- \mu \hn_j\nonumber\\
    &+\sum_\gamma \lambda_\gamma \hn_j (\hat{a}^\dagger_{j\gamma} + \hat{a}_{j\gamma}) + \omega_\gamma \hat{a}^\dagger_{j\gamma} \hat{a}_{j\gamma} \Big],
\end{align}
where $\hn_j=\hpsi^\dagger_j \hpsi_j - 1/2$. In the absence of dissipation, this model realizes the well-studied one-dimensional Mott transition between an LL and a MI at the half-filled density $\rho_0=1/2$. The baths are characterized by the spectral function $J(\omega)=\sum_\gamma \lambda_\gamma^2 \delta(\omega-\omega_\gamma)$ whose low-frequency behaviour is given by $J(\omega)= \alpha |\omega|^s$ for $|\omega| < 1/\tau_c$. The coupling $\alpha$ is the dissipation strength and the bath exponent $s>0$ defines the nature of the bath: sub-ohmic ($s<1$), ohmic ($s=1$) or super-ohmic ($s>1$) \cite{Caldeira_1983,Leggett_1987}.

We describe the fermionic degrees of freedom $\hpsi_j$ through bosonization about the half-filled density $\rho_0=1/2$ \cite{Haldane_1981_LL,Giamarchi,von_Delft_bosonisation,NdupuisCMUG2}. The density fluctuations are encoded in a bosonic field $\hphi(x)$ via
\begin{align}
    \label{eq:rho_phi}
    \hn_j = \hpsi^\dagger_j \hpsi_j - \frac12 \simeq -\frac{1}{\pi}\partial_x \hphi(x_j).
\end{align}
Particles added above half filling thus correspond to kinks of amplitude $-\pi$ in the bosonic field. In addition, the half-filled background allows for domain walls acting as ``half particles'' and represented by kinks of amplitude $-\pi/2$ (see Fig.~\ref{fig:model} (b)).

In the Euclidean (imaginary-time) action formalism, the bath operators $\hat{a}_{j\gamma}$ can be exactly integrated out \cite{Majumdar2023a,bouvdup_2025_bosonised1d}, leading to
\begin{align}\label{eq:phi_action}
    S=&\int_{x,\tau} \frac{1}{2 \pi K}\left[ u (\partial_x \phi)^2+ \frac{1}{u} (\partial_\tau \phi )^2\right] - g \cos(4\phi)\nonumber\\
    & +\frac{\mu}{\pi}\partial_x \phi - \alpha \underset{|\tau-\tau'|>\tau_c }{\int_{x,\tau,\tau'}} \frac{\cos(2\phi)\cos(2\phi')}{|\tau-\tau'|^{1+s}},
\end{align}
where $\int_x=\int_0^L\d x$, $\int_\tau=\int_0^\beta \d \tau$ and $\phi=\phi(x,\tau)$, $\phi'=\phi(x,\tau')$. On top of $\alpha$, $s$ and $\mu$, this action is parametrized by effective low-energy couplings: the Luttinger parameter $K$, the velocity $u$ and umklapp strength $g$. The dissipative term can be written as
\begin{equation}\label{eq:alpha_plus_minus}
    \cos(2\phi)\cos(2\phi')=\frac{\cos(2\phi+2\phi')}{2} + \frac{\cos(2\phi-2\phi')}{2}.
\end{equation}
The first term is similar to the umklapp coupling $g$ and the second penalizes temporal fluctuations of the field and is responsible for the DP discussed below.

\begin{figure}[h!]
    \centering
    \includegraphics[width=\linewidth]{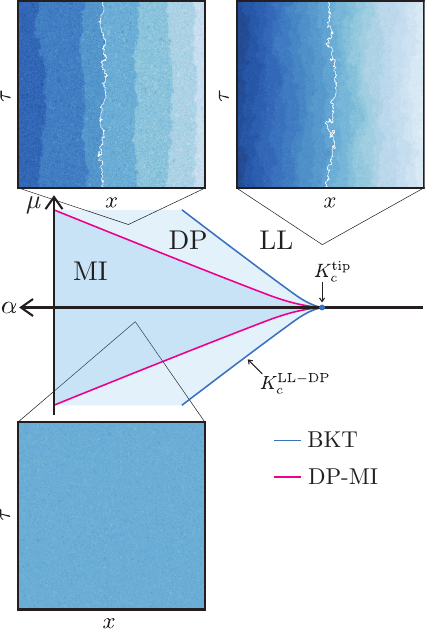}
    \caption{Center: Phase diagram inferred from the two limiting cases discussed in the main text. It is valid for $s<3/2$. For $s>3/2$, the DP disappears and the conventional LL--MI transition is recovered, with a BKT transition at $K_c^{\rm tip}$ at the tip and a second-order transition for $\mu\ne 0$. Blue lines and the blue dot denote BKT transitions, while magenta lines denote the DP--MI transition whose exponents depend on the bath exponent $s$. Sides: Typical field configurations $\phi(x,\tau)$ for the three phases for $s=1.25$ from a lattice realization of the model \ref{eq:phi_action} (for more details see \cite{SM}). The colormap encodes the value of the field $\phi$ with darker blue signifying higher values of $\phi$. Each kink has amplitude $\pi/2$ and traces the worldline of a half-particle (see Fig.~\ref{fig:model}(b)). One kink is highlighted in each of the DP and LL configurations.}
    \label{fig:phase_diagram}
\end{figure}

\paragraph{Phase diagram ---}
At commensurate filling $\rho=\rho_0$, the phase diagram of the model \eqref{eq:phi_action} is controlled by a LL--MI BKT transition at $K_c^{\rm tip}=\max(1/2,1-s/2)$, as shown in Refs.~\cite{Malatsetxebarria_2013,Bouverot_Dupuis_2024} and reviewed in the Supplemental Material~\cite{SM}. For the incommensurate regime $\rho\ne \rho_0$, Refs.~\cite{Malatsetxebarria_2013,Majumdar2023a,Ovni_2025} neglected the operators $g\cos(4\phi)$ and $\alpha\cos(2\phi+2\phi')$ which favour the MI. The remaining dissipative interaction $\alpha\cos(2\phi-2\phi')$ was found to drive a BKT transition at
\begin{equation}
    K_c^{\rm LL-DP}=1-\frac{s}{2}.
\end{equation}
It separates an LL at $K>K_c^{\rm LL-DP}$ from a DP at $K<K_c^{\rm LL-DP}$. Taken alone, this result suggests that the DP exists for all $s<2$, as depicted in Fig.~\ref{fig:phase_diagram}. Representative configurations of the three phases for $s=1.25$ are also shown. The MI contains no kinks, the DP displays dilute and temporally rigid kink worldlines, while their proliferation and roughening characterize the LL. The qualitative properties of the three phases are given in the End Matter.

However, we argue that this conclusion cannot hold throughout the entire range $s<2$ as it is also known that without dissipation ($\alpha=0$) the LL transitions to the MI for $K<K_c^{\rm LL-MI}=1/4$
\cite{Japaridze_1978,Pokrovsky_1979,Schulz_1980_CI,Papa_2001_sG}. Comparing $K_c^{\rm LL-DP}$ with $K_c^{\rm LL-MI}$, the DP can be reached before the MI only when
\begin{equation}
    K_c^{\rm LL-DP}>K_c^{\rm LL-MI},
\qquad\text{i.e.}\qquad s<\frac{3}{2}.
\end{equation}
Thus, the phase diagram in Fig.~\ref{fig:phase_diagram} is correct for $s<3/2$, while for $s>3/2$ the conventional direct LL--MI transition is recovered. Remarkably, the single-worldline theory developed below independently identifies the same threshold $s=3/2$, and further elucidates the universality class of the DP--MI transition.

\paragraph{Single-worldline theory ---}
\begin{figure}[h!]
    \centering
    \includegraphics[width=0.7\linewidth]{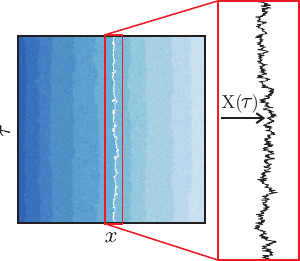}
    \caption{Near the DP--MI transition, the kinks become dilute. Neglecting microscopic overhangs, we parametrize one of them by a trajectory $X(\tau)$. The field configuration was obtained for $s=1$.}
    \label{fig:field_to_line}
\end{figure}
The DP--MI transition is accessible neither to the perturbative RG \cite{Giamarchi_1991}, since it occurs at strong-coupling, nor to the free-fermion mapping underlying the non-dissipative Mott transition \cite{Schulz_1980_CI}, since dissipation generates long-range interactions in imaginary time. We therefore introduce a different approach based on the geometry of the kink excitations.

As the system is tuned towards the MI, the kinks in the field $\phi$ become increasingly dilute. Right at the critical point, the critical properties should therefore be encoded in a field $\phi$ containing a single kink. If we neglect the small overhangs \footnote{We do not expect this approximation to have any impact on the universal properties of the kink line. See, for instance, \cite{Kolton_2023,Abitbol_2026}.}, the kink can be parametrized by a single-valued function $X(\tau)$ (see Fig.~\ref{fig:field_to_line}). As the kink's amplitude is $\pi/2$, we consider the ansatz
\begin{equation}
    \phi_X(x,\tau)= \frac{\pi}{2} \Theta(X(\tau)-x),
\end{equation}
with $\Theta$ the Heaviside function. Plugging this ansatz into the action Eq.~\eqref{eq:phi_action} leads to
\begin{align}\label{eq:S1_X}
    S[\phi_X]=&\int_{x,\tau} \frac{\pi}{8 K u} [\delta(X - x) \partial_\tau X]^2\nonumber\\
    &+ 2\alpha \underset{|\tau-\tau'|>\tau_c }{\int_{\tau,\tau'}} \frac{|X-X'|}{|\tau-\tau'|^{1+s}}.
\end{align}
The squared Dirac delta is just an artifact of the sharp-kink approximation $\phi(x,\tau)=\frac{\pi}{2} \Theta(X(\tau)-x)$. Using a smooth regularization of the step function (see End Matter) leads to
\begin{equation}
\label{eq:line_action}
S[X]=\frac12 \int_\tau (\partial_\tau X)^2 + \frac{\alpha}{2}\underset{|\tau-\tau'|>\tau_c}{\int_{\tau,\tau'}}\frac{|X-X'|}{|\tau-\tau'|^{1+s}},
\end{equation}
where we have absorbed unimportant constants by rescaling $X$ and $\alpha$. Equation~\eqref{eq:line_action} is the central result of this letter. It reduces the original $(1+1)$D quantum problem to a $(0+1)$D theory describing the fluctuations of a single critical worldline.

The roughness of the kink trajectory $X(\tau)$ directly determines the critical exponents. Since the critical point of the original model is invariant under the anisotropic rescaling $x,\tau \to bx, b^z \tau$ with $z$ the dynamical critical exponent, one expects
\begin{equation}
\label{eq:X_tau_scaling}
X(\tau)-X(0)
\sim
\tau^{1/z},
\end{equation}
which directly identifies the roughness exponent $\zeta=1/z$ of the kink line. It turns out that the other critical exponents $\nu$ and $\beta$ which characterize the vanishing of the doping $\delta \rho \sim |\mu-\mu_c|^\beta$ and the divergence of the correlation length $\xi \sim |\mu-\mu_c|^{-\nu}$ are also linked to $\zeta$ since they obey the scaling relations (see End Matter)
\begin{equation}\label{eq:beta_nu_z}
\beta=\nu=\frac{1}{z}=\zeta.
\end{equation}

\paragraph{Critical behaviour of the worldline ---}
Power counting around the Gaussian fixed point of the model \eqref{eq:line_action} gives the scaling dimension $[\alpha]=3/2-s$. This shows that for $s>3/2$, dissipation is irrelevant and the large-scale properties of $X(\tau)$ are that of the Gaussian free field, i.e. Brownian motion. There, $X(\tau) \sim \sqrt{\tau}$, hence $\zeta=1/2$. Using the scaling relations \eqref{eq:beta_nu_z}, this translates to $\beta=\nu=1/z=1/2$ which are the exponents of the generic Mott transition. This independently confirms the direct LL--MI transition, and thus the absence of the DP, for $s>3/2$.

For $s<3/2$, $\alpha$ becomes relevant. We expect it to rigidify the line $X(\tau)$, thereby reducing the roughness exponent $\zeta$. To determine $\zeta$, we use a Gaussian variational approach \cite{Giamarchi} which approximates the original action \eqref{eq:line_action} by a Gaussian action
\begin{equation}\label{eq:trial_action}
    S_0=\frac{1}{2}\int_\omega G^{-1}(\omega)|X(\omega)|^2.
\end{equation}
The optimal propagator $G(\omega)$ is found by minimizing a variational free energy (see \cite{SM}). We then extract $\zeta$ from the low-frequency behaviour $G^{-1}(\omega) \sim |\omega|^{1+2\zeta}$. This leads to
\begin{equation}\label{eq:zeta_var}
    \zeta=\begin{cases}
    1/2 &\text{ for } s>3/2,\\
    s-1 &\text{ for } 1<s<3/2,\\
    (s-1)/2 &\text{ for } s<1.
\end{cases}
\end{equation}
This confirms the Brownian behaviour for $s>3/2$ and establishes the intermediate positive roughness for $1<s<3/2$. The negative values of $\zeta$ for $s<1$ correspond to a flat interface. Therefore, the roughness exponent defined from $X(\tau)-X(0) \sim \tau^\zeta$ is $0$ for $s<1$.
\begin{figure}[h!]
    \centering
    \includegraphics[width=\linewidth]{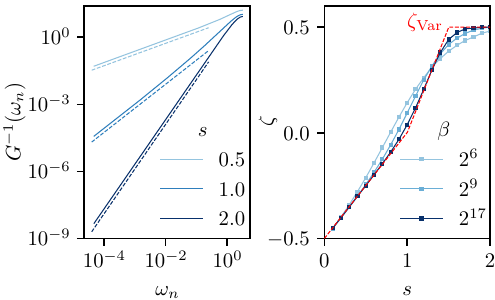}
    \caption{Left: Inverse propagator $G^{-1}(\omega_n)=1/\langle |X(\omega_n)|^2 \rangle$, with $\omega_n=2\pi n/\beta$, obtained from Monte Carlo simulations of Eq.~\eqref{eq:line_action} with $\beta=2^{17}$ (in units of the imaginary-time cutoff), $\alpha=1$ and varying $s$. Dashed lines show fits to the low-frequency power laws. Right: Roughness exponent $\zeta$ extracted from Monte Carlo simulations at $\alpha=1$ (blue points) compared to the variational result of Eq.~\eqref{eq:zeta_var} (red dashed line).}
    \label{fig:G_zeta}
\end{figure}

\paragraph{Numerical validation ---}
We test the single-worldline theory at two complementary levels. First, we simulate the single-worldline theory \eqref{eq:line_action} using the event-chain Monte Carlo algorithm \cite{Michel2014GenECMC,krauth2021ECMC,SM} and compute the propagator $G(\omega)=\langle |X(\omega)|^2 \rangle$ (Fig.~\ref{fig:G_zeta} left). Fitting the low-frequency behaviour $G^{-1}(\omega) \sim |\omega|^{1+2\zeta}$ gives the roughness exponent $\zeta$. The results are shown in Fig.~\ref{fig:G_zeta} (right) for different inverse temperatures $\beta$. As $\beta$ is increased, the data converge to the variational estimate, showing it is exact. The fact that the variational method gives an exact result for $\zeta$ is a hallmark of long-range interacting systems \cite{Sak_1973,Defenu2023}. Using $\zeta=1/z$ and Eq.~\eqref{eq:beta_nu_z}, the previous determination of $\zeta$ leads to
\begin{equation}\label{eq:exponents_var}
    \nu=\beta=\frac{1}{z}=\begin{cases}
    1/2 &\text{ for } s>3/2,\\
    s-1 &\text{ for } 1<s<3/2,\\
    0 &\text{ for } s<1,
\end{cases}
\end{equation}
which, for $s<1$, means the divergence of $\xi$ is slower than any power law.

\begin{figure}[h!]
    \centering
    \includegraphics[width=\linewidth]{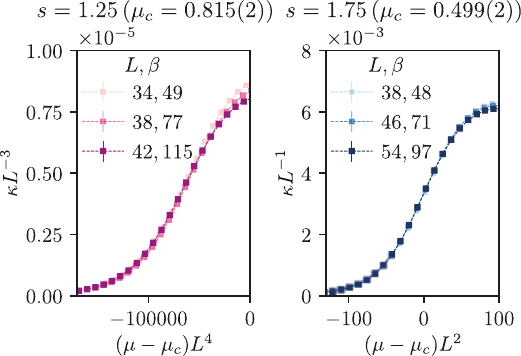}
    \caption{Finite-size scaling collapse for the compressibility $\kappa$ computed from the original $(1+1)$D model for $s=1.25$ (left) and $s=1.75$ (right). The inverse temperature is scaled as $\beta= 30^{1-z}L^z$ and $\mu_c$ is the only fitting parameter. The data were obtained using the single-histogram method \cite{Ferrenberg_1988}. Error bars were estimated by averaging over multiple independent runs.}
    \label{fig:kappa_collapse}
\end{figure}

Finally, we also test whether these predictions describe the original $(1+1)$D model \eqref{eq:phi_action}, without relying on the single-worldline theory, through a finite-size scaling study of the compressibility $\kappa= \lim_{q\to 0} (q/\pi)^2 \langle |\phi(q,0)|^2 \rangle$. From the scaling theory of second-order phase transitions \cite{Fisher_1989_SI,Wallin_1994_2DSI,NdupuisCMUG2} and using $\nu=1/z$, one expects the following finite-size scaling form
\begin{align}\label{eq:kappa_FSS}
    \kappa(L,\beta,\mu) &= L^{z-1} F\left((\mu-\mu_c) L^z,\beta/L^z\right),
\end{align}
with $F$ a scaling function. We numerically verify this prediction by simulating the $(1+1)$D model at fixed $\beta/L^z$ for $s=1.25$ and $s=1.75$ using, respectively, $z=4$ and $z=2$ as predicted by Eq.~\eqref{eq:exponents_var}. Figure~\ref{fig:kappa_collapse} shows the collapse of the compressibility with $\mu_c$ the single fitting parameter. The collapse is satisfactory, and we attribute its slight broadening to large finite-size effects which were already present in the $(0+1)$D model (see again the slow convergence in Fig.~\ref{fig:G_zeta} right).

\paragraph{Conclusion ---}
We have shown that dissipation splits the one-dimensional Mott transition for $s<3/2$, creating an intermediate DP between the LL and the MI, while for $s>3/2$ the conventional direct transition is recovered. The resulting DP--MI transition defines a new family of universality classes continuously indexed by the bath exponent $s$.

Although the DP--MI transition is accessible neither to perturbative RG nor to the free-fermion description of the conventional Mott transition, we showed that its critical behaviour follows from the geometry of a single kink worldline. The roughness of this worldline directly determines the critical exponents, as supported by numerical simulations of both the single-worldline theory and the original $(1+1)$D model. Candidate realizations of the dissipative transition include cold atoms coupled to transverse one-dimensional baths, mixed-dimensional mixtures \cite{LeBlanc_2007,Lamporesi_2010}, and shunted Josephson junctions \cite{Kuzmin2025}.

This dissipative phase diagram is reminiscent of the dirty-boson problem \cite{Giamarchi_1987,Giamarchi_1988,Fisher_1989_SI,Pollet_2009,Grison_2025}, where quenched disorder and the Bose glass (BG) phase replace the annealed disorder generated by the baths and the DP. However, the critical behaviour of the BG--MI transition remains unsettled \cite{Orignac_1999,Giamarchi_2001}. Whether our worldline approach could shed light on this transition is an interesting open question.

\paragraph{Acknowledgements ---}
We thank Thierry Giamarchi and Saptarshi Majumdar for fruitful discussions. O.B.-D. acknowledges the support of the French ANR under the grant ANR-22-CMAS-0001 (\emph{QuanTEdu-France} project). This work was supported by the French government through the France 2030 program (PhOM – Graduate School of Physics), under reference ANR-11-IDEX-0003 (Project Mascotte, L. Foini).

\bibliography{refs}

\onecolumngrid

\bigskip \bigskip
\centerline {\bf End Matter}
\bigskip

\twocolumngrid

\appendix

\section{Regularization procedure of the single-worldline theory}
To derive the single-worldline theory, it is necessary to regularize the sharp-kink ansatz used in the main text as
\begin{equation}
    \phi_X(x,\tau)=\Phi\left(\frac{X(\tau)-x}{a}\right),
\end{equation}
where $a$ is a microscopic length and $\Phi$ is a smooth function such that $\Phi(-\infty)=0$ and $\Phi(+\infty)=\pi/2$. A physically-motivated candidate for this function is $\Phi(x)=\arctan(\exp x)$ with $a=1/\Big[4\sqrt{\frac{\pi K}{u}(g+\frac{\alpha}{s{\tau_c}^s})}\,\Big]$ which is a solution of the equation of motion $\delta S/\delta \phi=0$. This replaces the squared Dirac delta $\delta^2(X-x)$ by $1/[\pi^2 a^2 \cosh^2(\frac{X-x}{a})]$ in Eq.~\eqref{eq:S1_X}. The previous expression is peaked around $x=X(\tau)$ and integrates to $2/(a\pi^2)$ so we can further approximate it to $\frac{2}{a\pi^2}\delta(X-x)$. Using this prescription, Eq.~\eqref{eq:S1_X} becomes
\begin{equation}
    S[\phi_X]=\frac{1}{4 \pi K a u} \int_\tau (\partial_\tau X)^2 + 2\alpha \!\!\underset{|\tau-\tau'|>\tau_c }{\int_{\tau,\tau'}} \frac{|X-X'|}{|\tau-\tau'|^{1+s}},
\end{equation}
which, after rescaling $X$ and $\alpha$, is Eq.~\eqref{eq:line_action} in the main text. We also note that the term $|X-X'|$ can be made analytic at the origin by using the smooth ansatz $\Phi$ but we refrain from doing so to keep the model as simple as possible.

\begin{figure*}[t!]
    \centering
    \includegraphics[width=\linewidth]{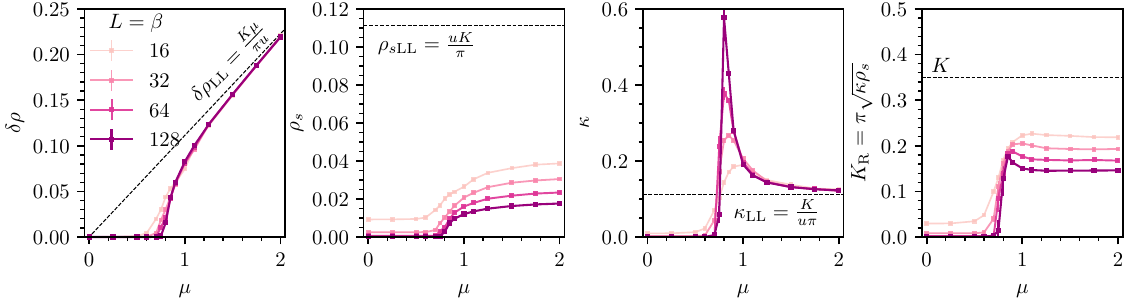}
    \includegraphics[width=\linewidth]{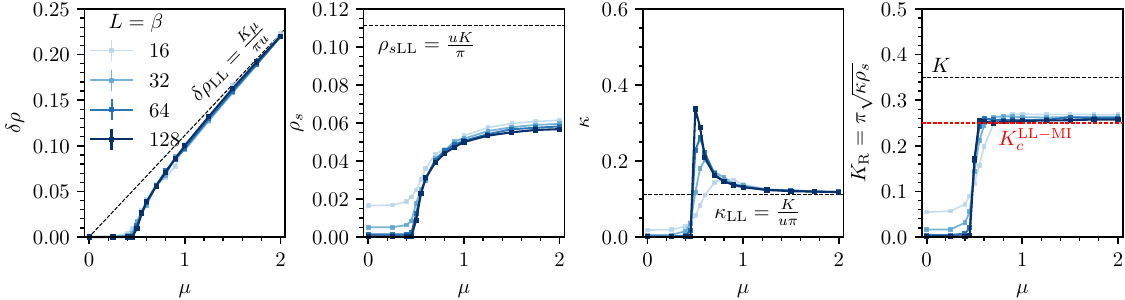}
    \caption{Characterization of the DP--MI transition for $s=1.25$ (top) and LL--MI transition $s=1.75$ (bottom) for increasing system sizes $L=\beta$. From left to right: doping $\delta\rho$, superfluid stiffness $\rho_s$, compressibility $\kappa$, and renormalized Luttinger parameter $K_{\rm R}$. In all panels, only $\mu$ is varied, with $K=0.35$, $u=1$, $g=0$, and $\alpha=0.5$. Black dashed lines indicate the corresponding values in the pure LL ($g=\alpha=0$), while the red dashed line in the lower-right panel marks $K_c^{\rm LL-MI}=1/4$. Error bars are often smaller than the symbols.}
    \label{fig:C_IC}
\end{figure*}

\section{Scaling relations at the DP--MI transition}
The DP--MI transition is characterized by the critical exponents $\beta$, $\nu$ and $z$ which describe the vanishing of the doping $\delta \rho$ and the divergence of the correlation length $\xi$ and correlation time $\xi_\tau$, i.e.
\begin{equation}
    \delta\rho \sim |\mu-\mu_c|^\beta, \quad \xi \sim |\mu-\mu_c|^{-\nu}, \quad \xi_\tau\sim\xi^z.
\end{equation}
Close to the transition, the doped excitations form a dilute gas and their typical separation, $1/\delta\rho$, is the only diverging length scale. It follows that $\xi \sim 1/\delta \rho$, so
\begin{equation}
    \beta=\nu.
\end{equation}
Next we relate $\nu$ to $z$ as follows. One first notices that $[\hat{H},\hat{N}]=0$ with $\hat{H}$ given in Eq.~\eqref{eq:H} and $\hat{N}=\sum_j \hpsi_j^\dagger \hpsi_j$ the total particle number. The eigenstates of $\hat{H}$ are thus
\begin{equation}\label{eq:GC_eigen}
    \hat{H}\ket{N,\gamma}=(E_{N,\gamma}-\mu N)\ket{N,\gamma},
\end{equation}
where $N$ is the particle number, $\gamma$ is the set of other quantum numbers, and $E_{N,\gamma}$ does not depend on $\mu$. Right at the transition, the Mott ground state $\ket{N=0,\rm MI}$ gets replaced by a new ground state $\ket{N=1,\gamma_0}$ with a single particle such that
\begin{align}
    \hat{H}\ket{0,\rm MI}=&E_{0,\rm MI}\ket{0, \rm MI},\\
    \hat{H}\ket{1,\gamma_0}=&(E_{1,\gamma_0}-\mu)\ket{1, \gamma_0}.
\end{align}
Coming from the MI, the gap is $\Delta=E_{1,\gamma_0}-\mu- E_{0,\rm MI}=\mu_c-\mu$ so the correlation time is $\xi_\tau \sim \Delta^{-1} = (\mu_c-\mu)^{-1}$.  From the definitions of $\nu$ and $z$, one expects $\xi_\tau \sim |\mu-\mu_c|^{-\nu z}$ so
\begin{equation}
    z\nu=1.
\end{equation}
Strictly speaking, we have shown that $z\nu=1$ holds on the MI side of the transition and not on the dissipative side. However, the exponents on both sides of a continuous phase transition are usually the same (except in some very pathological situations, see for instance \cite{Leonard_2015}).

\section{Numerical characterization of the phases}
We characterize the three phases through their compressibility $\kappa$, superfluid stiffness $\rho_s$ and doping $\rho_s$ defined as
\begin{align}
    \kappa &= \lim_{q\to 0}\frac{q^2}{\pi^2}\langle |\phi(q,0)|^2 \rangle\\
    \rho_s &= \lim_{\omega_n\to 0}\frac{\omega_n^2}{\pi^2}\langle |\phi(0,\omega_n)|^2 \rangle\\
    \delta \rho &= \rho-\rho_0=-\frac{1}{\pi}\langle \partial_x \phi \rangle.
\end{align}
We also consider the renormalized Luttinger parameter $K_{\rm R}=\pi\sqrt{\kappa\rho_s}$. The qualitative properties of the three phases are summarized in Table~\ref{tab}.

\begin{table}[h]
\centering
\begin{tabular}{||c c c c||}
\hline
& LL & DP & MI \\
\hline\hline
compressible ($\kappa$) & yes & yes & no \\
\hline
superfluid ($\rho_s$) & yes & no & no \\
\hline
doped ($\delta\rho$) & yes & yes & no \\
\hline
DC conducting & yes & if $s>1$ & no \\
\hline
\end{tabular}
\caption{Qualitative properties of the three phases.}
\label{tab}
\end{table}

Figure~\ref{fig:C_IC} shows these observables computed from Monte Carlo simulations of a lattice realization of the model \eqref{eq:phi_action} (see \cite{SM} for more details). For the compressibility and superfluid stiffness, we used the smallest possible momentum and frequency $q=2\pi/L$ and $\omega_1=2\pi/L$.

We first consider the case $s=1.25$ (Fig.~\ref{fig:C_IC} top) which we expect to exhibit the intermediate DP. For $\mu<\mu_c \simeq 0.8$, we clearly identify a MI with vanishing $\delta \rho$, $\kappa$ and $\rho_s$. For $\mu> \mu_c$, the MI transitions to a phase with non-zero doping $\delta \rho$ and compressibility $\kappa$. As the system size is increased the superfluid stiffness, and thus Luttinger parameter $K_\R$, seem to vanish. We identify this phase as the DP.

For $s=1.75$ (Fig.~\ref{fig:C_IC} bottom) both $\kappa$ and $\rho_s$ remain finite for $\mu >\mu_c\simeq 0.5$, identifying the phase as an LL. This implies that the DP found at $s=1.25$ has been replaced by an LL. Moreover, towards the transition, the renormalised Luttinger parameter $K_\R$ saturates at $K_c^{\rm LL-MI}=1/4$ which is characteristic of the non-dissipative Mott transition. These results independently confirm that the DP only exists for $s<3/2$.

\clearpage

\end{document}


\setcounter{equation}{0}
\setcounter{figure}{0}
\setcounter{table}{0}
\renewcommand{\theequation}{S\arabic{equation}}
\renewcommand{\theHequation}{S\arabic{equation}}
\renewcommand{\thefigure}{S\arabic{figure}}	
\renewcommand{\bibnumfmt}[1]{[S#1]}
\renewcommand{\citenumfont}[1]{S#1}

\graphicspath{{./figures/}}
	
\title{Supplemental Material: Single-worldline theory for a dissipative Mott transition}

\author{Oscar Bouverot-Dupuis}
\affiliation{Universit\'{e} Paris Saclay, CNRS, LPTMS, 91405, Orsay, France}
\affiliation{IPhT, CNRS, CEA, Universit\'{e} Paris Saclay, 91191 Gif-sur-Yvette, France}

\author{Alberto Rosso}
\affiliation{Universit\'{e} Paris Saclay, CNRS, LPTMS, 91405, Orsay, France}

\author{Laura Foini}
\affiliation{IPhT, CNRS, CEA, Universit\'{e} Paris Saclay, 91191 Gif-sur-Yvette, France}
	
\maketitle
	
\onecolumngrid
	
In this Supplemental Material, we first review the perturbative arguments which led to believe that the DP existed for all $s<2$ in Sec.~\ref{SM:pert_analysis}. This is based on perturbative renormalization group (RG) computations for the commensurable and incommensurable cases, which we rederive in a unified setting in Sec.~\ref{SM:RG}. Section~\ref{SM:variational} then details the variational computation of the roughness exponent $\zeta$ of the single-worldline theory. Finally, we present a rapid numerical study of the LL--DP transition in Sec.~\ref{SM:LL_DP_BKT} and detail our Monte Carlo algorithms for the original $(1+1)$D model and the single-worldline theory in Secs.~\ref{SM:MC_phi} and \ref{SM:MC_X}.
	
\tableofcontents

\section{Perturbative analysis}
\label{SM:pert_analysis}
In order to analyse the $(1+1)$D model
\begin{align}\label{eq:phi_action}
    S=&\int_{x,\tau} \frac{1}{2 \pi K}\left[ u (\partial_x \phi)^2+ \frac{1}{u} (\partial_\tau \phi )^2\right] - g \cos(4\phi) +\frac{\mu}{\pi}\partial_x \phi - \alpha \underset{|\tau-\tau'|>\tau_c }{\int_{x,\tau,\tau'}} \frac{\cos(2\phi)\cos(2\phi')}{|\tau-\tau'|^{1+s}},
\end{align}
we perform a renormalisation group (RG) study perturbative in $\alpha$ and $g$. This is done in the canonical ensemble where we trade the chemical potential $\mu$ for a fixed microscopic doping $\delta \rho= -\langle \partial_x \phi \rangle/\pi$. After a renormalisation ``time" $l$, this procedure yields renormalised couplings $\alpha(l)$, $g(l), K(l), u(l)$ which account for the fluctuations between the microscopic scales $a$, $\tau_c$ and larger ones $a(l)=ae^l$, $\tau_c(l)=a(l)/u(l)$. A convenient set of variables to express the RG flow is given by the dimensionless couplings $\tilde g(l)= 2\pi g(l) a(l) \tau_c(l)$, $\tilde \alpha (l) = 2\pi \alpha(l) a(l) \tau_c(l)^{1-s}$, and $\tilde {\delta \rho} = \delta \rho\, a(l)=\tilde {\delta \rho} (0) e^l$
for which the RG equations read (see Sec.~\ref{SM:RG} for more details)
\begin{align}
    \label{eq:RG_K}
    \frac{\d}{\d l}&\frac{1}{K}=\tilde g^2 J_0(4 \pi \tilde {\delta\rho})+ \tilde \alpha,\\
    \label{eq:RG_u}
    \frac{\d}{\d l}& u=-uK\left(\tilde g^2 J_2(4 \pi \tilde {\delta\rho}) + \tilde \alpha\right),\\
    \label{eq:RG_g}
    \frac{\d}{\d l}&\tilde g=(2-4K) \tilde g+\tilde \alpha,\\
    \label{eq:RG_alpha}
    \frac{\d}{\d l}& \tilde \alpha=(2-s-2K) \tilde \alpha + \tilde g \tilde \alpha J_0(3 \pi \tilde {\delta\rho}).
\end{align}
The $J_n(z)$s are Bessel functions which are strongly suppressed for $z\gg 1$ and are such that $J_0(0)=1$, $J_1(0)=J_2(0)=0$ and $J_0(\infty)=J_1(\infty)=J_2(\infty)=0$. The fact they are oscillating functions is a non-universal feature due to the sharp real-space cutoff $a$ used in the RG procedure.

To make sense of these RG equations, we first notice that the dimensionless doping $\tilde {\delta\rho}(l)=\delta \rho \, a e^l$ remains zero if $\delta \rho=0$, and diverges to $\pm \infty$ for large $l$ otherwise. This distinguishes the commensurate transition at $\delta \rho=0$, from the incommensurate one at $\delta \rho \ne 0$. If we consider a transition from an LL, since the doping is related to the chemical potential $\mu$ as $\delta \rho=K\mu/(\pi u)$, this differentiates the transition at $\mu=0$ (the tip of the lobe in Fig.~(2) of the main text) from those at $\mu \ne 0$ (the sides of the lobe in Fig.~(2) of the main text). We now analyse separately these two cases.

\subsection{Commensurate transition}
If $\delta \rho=0$, we are considering the commensurate transition where one approaches the Mott lobe from its tip. In this case, setting $\tilde {\delta \rho}(l)=0$ in the RG equations yields
\begin{align}
    \label{eq:RG_K1}
    \frac{\d}{\d l}&\frac{1}{K}=\tilde g^2+ \tilde \alpha,\\
    \label{eq:RG_u1}
    \frac{\d}{\d l}& u=-uK \tilde \alpha,\\
    \label{eq:RG_g1}
    \frac{\d}{\d l}&\tilde g=(2-4K) \tilde g+\tilde \alpha,\\
    \label{eq:RG_alpha1}
    \frac{\d}{\d l}& \tilde \alpha=(2-s-2K) \tilde \alpha + \tilde g \tilde \alpha.
\end{align}
These equations were already derived in Ref.~\cite{Malatsetxebarria_2013,Bouverot_Dupuis_2024} by directly setting $\mu=0$ in Eq.~\eqref{eq:phi_action}. The key takeaway is that the flow departs from the LL as soon as $\tilde g$ or $\tilde \alpha$ flow to strong coupling. Looking at Eqs.~\eqref{eq:RG_g1} and \eqref{eq:RG_alpha1}, this happens for $K<K_c^{\rm tip}$ with
\begin{equation}
    K_c^{\rm tip}=\max(1/2,1-s/2),
\end{equation}
and the flow near the critical point $K_c^{\rm tip}$ is of the BKT form. For $K>K_c^{\rm tip}$, the couplings $\tilde \alpha$, $\tilde g$ eventually renormalise to $0$ and the low-energy physics is described by the LL action
\begin{equation}\label{eq:LL_action}
    S_{\rm LL}=\int_{q,\omega} \frac{u_\R q^2 + \omega^2/u_\R}{2 \pi K_\R} |\phi(q,\omega)|^2,
\end{equation}
where the renormalised couplings are $u_\R=u(l=\infty)$ and $K_\R=K(l=\infty)$, and $\phi(q,\omega)=\int_{x,\tau} e^{-i(qx+\omega \tau)}\phi(x,\tau)$ with $\int_{q,\omega}=\int \frac{\d q \d \omega}{4\pi^2}$. On the contrary, when $K<K_c^{\rm tip}$ both $\tilde \alpha$ and $\tilde g$ grow large and confine the field to a minimum, e.g. $\phi \simeq 0$. This justifies expanding the action \eqref{eq:phi_action} at quadratic order in $\phi$ such that
\begin{align}\label{eq:MI_action}
    S_{\rm MI}\! =\! \int_{q,\omega} \! \left[\frac{u_\R q^2 + \omega^2/u_\R }{2\pi K_\R} + 8 g_\R + \frac{8\alpha_\R}{s{\tau_c}^s} \right] \! |\phi(q,\omega)|^2.
\end{align}
This gapped phase is a MI. With the ansatzes (\ref{eq:LL_action},\ref{eq:MI_action}), one finds that the compressibility$\kappa$, superfluid stiffness $\rho_s$ (see End matter for their definitions) and conductivity (in units of $e^2/\hbar$)
\begin{align}\label{eq:sigma_def}
    \sigma(\omega)=&{\rm Re}\left\{\frac{1}{\pi^2} \left[\omega_n \langle|\phi(\omega_n)|^2\rangle\right]_{i\omega_n \to \omega + i\delta} \right\},
\end{align}
are given by
\begin{equation}
    \kappa_{\rm LL}= \frac{K_\R}{\pi u_\R}, \quad {\rho_s}_{\rm LL}=\frac{K_\R u_\R}{\pi}, \quad \sigma(\omega)=K_\R u_\R \delta(\omega),
\end{equation}
in the LL, and
\begin{equation}
     \kappa_{\rm MI} = {\rho_s}_{\rm MI} = \sigma(\omega) = 0,
\end{equation}
in the MI. From these observables one can also extract the renormalised Luttinger parameter $K_\R=\pi \sqrt{\kappa \rho_s}$ in the LL.

\subsection{Incommensurate transition}

In the incommensurate case where $\delta \rho \ne 0$, we are considering the sides of the Mott lobe. This corresponds to $\tilde {\delta \rho}(l)$ diverging to $\pm \infty$. Introducing the tilt of the field $\phi$ as $\delta=-\delta \rho/\pi$, it is then convenient to define $\varphi(x,\tau)=\phi(x,\tau)-\delta x$ which is not tilted. Since the umklapp term $\cos(4\phi)= \cos(4\varphi + 4\delta x)$ oscillates, it averages out at large scales and is simply discarded \cite{Giamarchi}. Similarly, retaining only the non-oscillating part of the dissipative term replaces $\cos(2\phi)\cos(2\phi')$ by $\cos(2\varphi-2\varphi')/2$. This yields the effective action
\begin{align}\label{eq:phi_action_incom}
    S=&\int_{x,\tau} \frac{1}{2 \pi K}\left[ u(\partial_x \varphi)^2+\frac{1}{u}(\partial_\tau \varphi)^2\right] - \frac{\alpha}{2} \underset{|\tau-\tau'|>\tau_c }{\int_{x,\tau,\tau'}} \frac{\cos(2\varphi-2\varphi')}{|\tau-\tau'|^{1+s}}.
\end{align}
To understand the phases of this model we make use of the RG equations Eqs.~(\ref{eq:RG_K}-\ref{eq:RG_alpha}) which, for $\tilde {\delta \rho}= \pm\infty$, reduce to
\begin{align}
    \label{eq:RG_K2}
    \frac{\d}{\d l}&\frac{1}{K}= \tilde \alpha,\\
    \label{eq:RG_u2}
    \frac{\d}{\d l}& u=-uK \tilde \alpha,\\
    \label{eq:RG_alpha2}
    \frac{\d}{\d l}& \tilde \alpha=(2-s-2K) \tilde \alpha,
\end{align}
where we have omitted the RG equation for $\tilde g$ since it decouples from the others. These equations were studied in \cite{Malatsetxebarria_2013,Majumdar2023a,Ovni_2025} and describe a BKT transition at
\begin{equation}\label{eq:K_c_LL_DP}
    K_c^{\rm LL-DP}=1-s/2.
\end{equation}
For $K>K_c^{\rm LL-DP}$, $\tilde \alpha$ is irrelevant from Eq.~\eqref{eq:RG_alpha2} and the low-energy limit of Eq.~\eqref{eq:phi_action_incom} is the LL action Eq.~\eqref{eq:LL_action} with $\varphi$ instead of $\phi$. For $K<K_c^{\rm LL-DP}$, $\tilde \alpha$ becomes relevant. Performing a quadratic expansion of Eq.~\eqref{eq:phi_action_incom} and keeping the dominant low-frequency behaviour yields
\begin{align}
    S_{\rm DP}=&\int_{q,\omega}  \left[\frac{u_\R q^2}{2\pi K_\R}+ \alpha_\R I(s) |\omega|^s \right]|\varphi(q,\omega)|^2,
\end{align}
with $I(s)= -4\cos(\pi s/2)\Gamma(-s)>0$. This gapless action scales space and time as $\omega=q^z$ with the dynamical exponent $z=2/s$ which clearly differentiates it from the LL where $z=1$. The fluctuations in space and time of $\varphi$ are furthermore bounded, indicating long-range order corresponding to a density wave \cite{Majumdar2023a}. The compressibility, superfluid stiffness and conductivity of this DP are found to be
\begin{equation}
    \kappa_{\rm DP} = \frac{K_\R}{\pi u_\R}, \quad {\rho_s}_{\rm DP}=0, \quad \sigma(\omega) = \frac{\sin(\pi s/2)}{2 \pi^2 \alpha_\R I(s)}\omega^{1-s}.
\end{equation}
The DC conductivity $\sigma(\omega\to 0)$ depends on $s$:  for $s>1$ it is infinite but less divergent than the LL which has a Drude peak, for $s=1$ it is finite as for an ohmic conductor, while for $s<1$ it vanishes and the DP becomes insulating.

\subsection{Putative phase diagram}

The preceding perturbative analysis suggests the phase diagram shown in Fig.~(2) of the main text, with the DP existing as long as $K_c^{\rm LL-DP}>0$, i.e. $s<2$. However, our single-worldline theory shows that this conclusion is incorrect: the DP exists only for $s<3/2$. The origin of this discrepancy is that the perturbative analysis assumes that entering the incommensurate regime merely requires $\delta\rho \ne 0$. However, as shown in the main text, the incommensurate regime is characterized by both $\delta\rho \ne 0$ and $K>1/4$. Consequently, Eq.~\eqref{eq:phi_action_incom} is only valid for $K>1/4$. Imposing this additional constraint restricts the putative DP and yields the correct threshold $s<3/2$.

\section{Perturbative RG equations}
\label{SM:RG}

This section derives the perturbative RG equations. As highlighted in Ref.~\cite{Horovitz_1983_CI_RG} which studied the non-dissipative case, a renormalisation procedure is only possible if we work in the canonical ensemble where we trade the constant chemical potential for a constant doping $\delta\rho=-\frac{1}{\pi}\partial_x \phi$. This indeed allows to introduce a non-tilted field $\varphi(x,\tau)=\phi(x,\tau) - \delta x$, $\delta = - \pi \delta \rho$, which possesses well-defined Fourier modes and can thus be coarse grained. The action for this new field is $S=S_{\rm LL}+ S_g + S_\alpha$ with
\begin{align}
    S_{\rm LL}=&\int_{x,\tau}\frac{1}{2\pi K} \left[u(\partial_x \varphi )^2 +\frac{1}{u}( \partial_\tau \varphi)^2\right]\\
    S_g=&- g \int_{x,\tau} \cos(4\varphi+4 \delta x)\\
    S_\alpha=& - \alpha \underset{|\tau-\tau'|>\tau_c}{\int_{x,\tau,\tau'}}\frac{\cos(2 \varphi + 2 \delta x)\cos(2 \varphi' + 2 \delta x)}{|\tau-\tau'|^{1+s}},
\end{align}
and is the starting point of our RG computation. We decide to derive the RG equations perturbatively around the Gaussian action $S_{\rm LL}$.

To perform our perturbative renormalisation group computation, we use the language of the operator product expansion as in \cite{cardy_RG}. We will track the one-loop corrections, which means going to order $\mathcal{O}(g^2,\alpha)$ for the RG equations of $K$, $u$, $g$ and $\mathcal{O}(\alpha g)$ for that of $\alpha$. To do so, we expand the partition function up to order $\mathcal{O}(\alpha,g^2,\alpha g)$ such that
\begin{align}\label{eq:action_expansion}
    Z=\int \mathcal{D}\varphi\, e^{-S_{\rm LL}[\varphi]-S_g[\varphi]-S_\alpha[\varphi]} = Z_{\rm LL}\Big[1- \langle S_g\rangle(a)-\langle S_\alpha\rangle(a) +\frac{\langle S_g^2\rangle(a)}{2}+\langle S_\alpha S_g\rangle(a) \Big],
\end{align}
where the dependency on the cutoff $a$ has been made explicit and $\langle . \rangle$ refers to the average with respect to the LL action. We then perform a rescaling $a\to a'=a(1+ \d l)$ (so $\tau_c \to \tau_c'=\tau_c(1+\d l)u/u'$) and ask how the couplings should vary to preserve the partition function $Z$. $Z_{\rm LL}$ being an RG fixed-point, we only need to consider the variations of the averaged interacting terms. Since this is a coarse graining procedure, the variation of the couplings naturally gives the RG equations.

\paragraph{Operator product expansions}
In practice, when rescaling $a\to a'$, nearby operators merge into new operators. This process is captured by the operator product expansion (OPE) \cite{Wiese_2000} so we first derive the important OPEs.
\begin{align}
    :\cos [2p(\varphi(r)+\delta x)]::\cos [2p(\varphi(r')+\delta x')]: =&\frac{a^{2p^2K}}{2}:\cos[4p(\varphi(R)+\delta X)]: - p^2\frac{:(\hat r\cdot \nabla \varphi(R))^2):}{|\hat r|^{2p^2K}}\cos(2p\delta \hat x)+...,\! \label{eq:OPE_1}\\
    :\cos[4(\varphi(r)+\delta x)]::\cos[2(\varphi(r')+\delta x')]:=&\frac{:\cos[2(\varphi(R)+\delta X)]:}{2a^{4K}}\cos (3\delta \hat x)+...,\label{eq:OPE_2}
\end{align}
where $:\cdot:$ denotes bosonic normal ordering and $r=(x,u\tau)$, $r'=(x',u\tau')$, $\hat r=r'-r$, $R=(r+r')/2$. The dots $\dots$ indicate that we discard irrelevant operators. These relations are proven using the identities $\langle \varphi(r)\varphi(0)\rangle = -\frac{K}{2}\log(|r|+a)$ and $:e^{i2p\varphi(r)}:=\frac{e^{i2p\varphi(r)}}{\langle e^{i2p\varphi(r)} \rangle}=\frac{e^{i2p\varphi(r)}}{a^{p^2 K}}$  where $a$ is the UV cutoff. 

\paragraph{Normal ordering} Under the rescaling $a \to a'$, the average of a vertex operator $\langle e^{i\sum_j \varphi(x_j,\tau_j)} \rangle$ is not invariant and renormalises according to its scaling dimension. To extract this rescaling, it is convenient to systematically normal order all vertex operators as
\begin{equation}
    :e^{i\sum_j \varphi(x_j,\tau_j)}: = e^{i\sum_j \varphi(x_j,\tau_j)}/\langle e^{i\sum_j \varphi(x_j,\tau_j)} \rangle,
\end{equation}
since, clearly, $\langle :e^{i\sum_j \varphi(x_j,\tau_j)}:\rangle=1$ and does not renormalise. Normal ordering the interactions in the original action gives
\begin{align}
    S_g=&-\frac{g}{a^{-4K}}\int_{x,\tau} :\cos (4\varphi + 4\delta x ):,\\
    S_\alpha=&-\frac{\alpha }{a^{-2K}}\hspace{-0.1cm}\underset{|\tau-\tau'|>\tau_c}{\int_{x,\tau,\tau'} } \hspace{-0.6cm} \frac{:\cos ( 2 \varphi + 2\delta x )::\cos ( 2 \varphi' + 2\delta x):}{|\tau-\tau'|^{1+s}}.
\end{align}

\paragraph{The computation}
After the infinitesimal rescaling, the new couplings are denoted by
\begin{align}
    \delta'= \delta+\d \delta, \quad g'=g+\d g, \quad \alpha'=\alpha+\d \alpha, \quad K'=K+\d K, \quad u'=u+\d u. 
\end{align}
In the following, we assume that $\d K,\d u =\mathcal{O}(g^2,\alpha)$ (which will be checked in the end) so as not to write unnecessary lengthy equations. Moreover, since $\delta$ is related to the field tilt which is macroscopically kept fixed, we also have $\d \delta=0$. Keeping only the first order in coupling variations, $\langle S_g\rangle(a')$ becomes
\begin{align}
    \langle S_g\rangle (a')=&\langle S_g\rangle (a)\left[ 1+\frac{\d g}{g} + 4K\d l\right].
\end{align}
The next term $\langle S_\alpha\rangle (a')$ requires a bit more work. We start by splitting the imaginary-time integral as $\int_{|\tau-\tau'|>\tau_c'} =\int_{|\tau-\tau'|>\tau_c} - \int_{\tau_c'>|\tau-\tau'|>\tau_c}$. The first term is straightforward to compute, while the second, which involves only fields at very close positions, can be evaluated using the OPE \eqref{eq:OPE_1}. The sum of the two terms yields
\begin{align}
    \langle S_\alpha\rangle(a')=&\langle S_\alpha\rangle (a)\left[ 1+\frac{\d\alpha}{\alpha} + 2K\d l\right] - \langle S_g \rangle (a)\frac{\alpha}{g \tau_c^s}\d l - 2\alpha {\tau_c}^{2-s}\int_{x,\tau} \langle:(\partial_\tau \varphi)^2:\rangle \d l.
\end{align}
The third term, namely $\frac{1}{2}\langle S_g^2\rangle(a')$, reads
\begin{align}
    \frac{\langle S_g^2\rangle(a')}{2}=&\frac{g'^2}{2 u'^2 a'^{-8K'}}\underset{|r-r'|>a'}{\int \d^2 r \d^2 r'}\langle :\cos (4\varphi+4\delta x)::\cos (4\varphi'+4\delta x'):\rangle,
\end{align}
where $\varphi'=\varphi(r')$ and the condition $|r-r'|>a'$ comes from the whole theory being regulated at the scale $a$. For this term, one again splits the limits of the integral into $\int_{|r-r'|>a} - \int_{a'>|r-r'|>a}.$ The first part is easily tractable while the second requires the use of the OPE \eqref{eq:OPE_1}. This yields
\begin{align}
    \frac{\langle S_g^2\rangle(a')}{2}=&\frac{\langle S_g^2\rangle(a)}{2}\left[1+\frac{2\d g}{g} + 8K\d l \right]\\
    &+ 2g^2 a^2 \tau_c^2\int_0^{2\pi}\d\theta \cos(4 \delta a \cos \theta)\int_{x,\tau} \left[u\langle:(\partial_x\varphi)^2:\rangle\cos^2\theta+\frac{\langle:(\partial_\tau\varphi)^2:\rangle}{u}\sin^2\theta\right]\d l.\nonumber
\end{align}
The last term $\langle S_g S_\alpha \rangle(a')$ is
\begin{align}
    \langle S_g S_\alpha\rangle(a')=& \frac{g' \alpha' }{a'^{-6K'}u'^3}\int_{\substack{|r-r'|>a'\\|r'-r''|>a'\\|r-r''|>a'}}\frac{\delta(x'-x'')}{|\tau'-\tau''|^{1+s}}\langle :\cos (4\varphi+4\delta x)::\cos ( 2\varphi' + 2\delta x')::\cos ( 2\varphi'' + 2\delta x''):\rangle.
\end{align}
The integral is then separated as (at order $\mathcal{O}(\d l)$)
\begin{equation}
    \int_{\substack{|r-r'|>a'\\|r'-r''|>a'\\|r-r''|>a'}}=\int_{\substack{|r-r'|>a\\|r'-r''|>a\\|r-r''|>a}} - \int_{\substack{a>|r-r'|>a'\\|r'-r''|>a\\|r-r''|>a}}  - \int_{\substack{|r-r'|>a\\|r'-r''|>a\\a>|r-r''|>a'}} - \int_{\substack{|r-r'|>a\\a>|r'-r''|>a'\\|r-r''|>a}}.
\end{equation}
The first contribution simply gives the scaling dimension of the operator, the second and third are equal and renormalise $\alpha$, and the last one renormalises $g$ at a higher order than what we are looking for, so we drop it. Using the OPE \eqref{eq:OPE_2}, one shows that retaining the relevant terms leads to
\begin{align}
    \langle S_g S_\alpha\rangle(a')=&\langle S_g S_\alpha\rangle(a)\Big[1+\frac{\d g}{g}+\frac{\d\alpha}{\alpha} + 6K\d l\Big]+\langle S_\alpha \rangle(a) g a \tau_c \int_0^{2\pi}\d\theta \cos(3\delta a \cos\theta)\d l.
\end{align}
Putting everything together, one arrives at
\begin{align}
    &\left[ \langle S_g\rangle+\langle S_\alpha\rangle-\frac{\langle S_g^2\rangle}{2}-\langle S_g S_\alpha\rangle \right](a')-\left[ \langle S_g\rangle+\langle S_\alpha\rangle-\frac{\langle S_g^2\rangle}{2}-\langle S_g S_\alpha\rangle \right](a)\nonumber\\
    =&\langle S_g\rangle (a)\left[ \frac{\d g}{g} + 4K\d l-\frac{\alpha}{g \tau_c^s}\d l\right]\nonumber\\
    &+\langle S_\alpha\rangle (a)\left[ \frac{\d\alpha}{\alpha} + 2K\d l - 2\pi g a \tau_c J_0(3\delta a)\d l\right]\nonumber\\
    &-\langle S_g^2\rangle(a)\left[\frac{\d g}{g} + 4K\d l \right]\nonumber\\
    &-\langle S_g S_\alpha\rangle(a)\Big[\frac{\d g}{g}+\frac{\d\alpha}{\alpha}+6K\d l\Big].\nonumber\\
    &- 2 g^2 a^2 \tau_c^2 \pi \int_{x,\tau} \left[u\langle:(\partial_x\varphi(r))^2:\rangle (J_0(4\delta a)-J_2(4\delta a))+\frac{\langle:(\partial_\tau\varphi(r))^2:\rangle}{u}(J_0(4\delta a)+J_2(4\delta a))\right]\d l\nonumber\\
    &-2\alpha \tau_c^{2-s}\int_{x,\tau}\langle:(\partial_\tau \varphi(x,\tau))^2:\rangle \d l,
\end{align}
where $J_n(x)=\frac{1}{\pi}\int_0^\pi \d\theta\cos(n\theta-x\sin \theta)$ is the $n$-th Bessel function. Upon imposing that the partition function $Z$ remains unchanged, the RG equations for $\alpha$ and $g$ can be directly read off. Those for $K$ and $u$ are found by re-exponentiating the remaining quadratic terms and compensating them by an appropriate change of $K$ and $u$ in $S_{\rm LL}$. Using the dimensionless couplings $\tilde g = 2 \pi g a \tau_c$, $\tilde \alpha = 2 \pi \alpha a \tau_c^{1-s}$ and $\tilde {\delta \rho} = - \delta a/\pi$, one ends up with
\begin{align}\label{eq:RG_K_canonical}
    \frac{\d}{\d l}&\frac{1}{K}=\tilde g^2 J_0(4 \pi \tilde {\delta \rho})+ \tilde \alpha,\\
    \frac{\d}{\d l}& u=-uK\left(\tilde g^2 J_2(4 \pi \tilde {\delta \rho}) + \tilde \alpha\right),\\
    \frac{\d}{\d l}&\tilde g=(2-4K) \tilde g+\tilde \alpha,\\
    \label{eq:RG_alpha_canonical}
    \frac{\d}{\d l}& \tilde \alpha=(2-s-2K) \tilde \alpha + \tilde g \tilde \alpha J_0(3 \pi \tilde {\delta \rho}),\\
    \frac{\d}{\d l}&\tilde {\delta \rho} = \tilde {\delta \rho},
\end{align}
which are the RG equations given in the main text.

\section{Variational method for the single-worldline theory}
\label{SM:variational}

This section details the variational computation of the roughness exponent $\zeta$ of the single-worldline theory. We follow closely the method outlined in \cite{Giamarchi}. We wish to approximate the action
\begin{equation}\label{eq:action_line_app}
    S=\frac{1}{2}\int_\tau (\partial_\tau X)^2 + \frac{\alpha}{2} \underset{|\tau-\tau'|>\tau_c}{\int_{\tau,\tau'}} \frac{|X'-X|}{|\tau-\tau'|^{1+s}},
\end{equation}
using a Gaussian trial action
\begin{equation}\label{eq:action_var_app}
    S_0=\frac{1}{2\beta}\sum_{\omega_n \ne 0} G^{-1}(\omega_n)|X(\omega_n)|^2,
\end{equation}
with $X(\tau)=1/\beta\sum_{\omega_n} e^{-i\omega_n \tau}X(\omega_n)$ and $G(\omega_n)=G(-\omega_n)$. We have excluded the zero-frequency mode since it decouples completely from the action \eqref{eq:action_line_app}. The optimal propagator $G(\omega_n)$ is defined as that which minimises the variational free energy
\begin{equation}
    F_{\rm var}=-T\log(Z_0) + T\langle S - S_0\rangle_0,
\end{equation}
with $Z_0$ and $\langle \cdot \rangle_0$ the partition function and average associated to \eqref{eq:action_var_app}. $F_{\rm var}$ is indeed bounded from below by the free energy $F=-T\log(Z)$ with $Z$ the partition function of \eqref{eq:action_line_app}, since
\begin{align}
    F&=-T\log(Z_0\langle e^{-S+S_0}\rangle_0) \le -T\log(Z_0) + T\langle S - S_0\rangle_0=F_{\rm var}.
\end{align}
To compute the variational free energy, first observe that the last term $\langle S_0 \rangle_0 \sim \sum_{\omega_n} G^{-1}(\omega_n) G(\omega_n)$ is constant and can be dropped. Next, being careful not to overcount frequencies, one finds
\begin{align}
    -T\log(Z_0)=-\frac{T}{2}\sum_{\omega_n\ne 0} \log(G(\omega_n)).
\end{align}
The term $T\langle S \rangle_0$ can be rewritten as
\begin{equation}
    T\langle S \rangle_0 = \frac{T}{2}\sum_{\omega_n \ne 0} \omega_n^2 G(\omega_n) + T \frac{\alpha}{2} \underset{|\tau-\tau'|>\tau_c}{\int_{\tau,\tau'}} \hspace{-0.2cm} \frac{\langle |X'-X| \rangle_0}{|\tau-\tau'|^{1+s}}.
\end{equation}
The average of the absolute value is computed as follows.
\begin{align}
    \langle |X(\tau)-X(\tau')| \rangle_0=&\left\langle \int_{-\infty}^{+\infty} \d k \frac{1 - \cos[k(X(\tau)-X(\tau'))]}{\pi k^2} \right\rangle_0\nonumber\\
    =&\int_{-\infty}^{+\infty} \d k \frac{1 - e^{- k^2/\beta \sum_{\omega_n \ne 0} G(\omega_n)\big[1-\cos(\omega_n(\tau-\tau'))\big]}}{\pi k^2}\nonumber\\
    =&\frac{2}{\sqrt{\pi}}\sqrt{\frac{1}{\beta} \sum_{\omega_n \ne 0} G(\omega_n) \big[1-\cos(\omega_n(\tau-\tau'))\big]}
\end{align}
Putting everything together yields the variational action which, in the $\beta \to \infty$ limit, reads
\begin{align}
    &F_{\rm var}=-\frac{1}{2} \int_\omega \log(G(\omega)) + \frac{1}{2}\int_\omega \omega^2 G(\omega) + \frac{\alpha}{\sqrt{\pi}}\underset{|\tau|>\tau_c}{\int_\tau} \frac{1}{|\tau|^{1+s}} \sqrt{\int_\omega G(\omega) [1-\cos(\omega\tau)] }.
\end{align}
The optimal $G(\omega)$ is defined such that
\begin{align}
    \frac{\delta F_{\rm var}}{\delta G(\omega)}[G]=0,
\end{align}
and thus satisfies
\begin{align}\label{eq:var_opti}
    G^{-1}(\omega) = \omega^2 + \frac{\alpha}{\sqrt{\pi}}\underset{|\tau|>\tau_c}{\int_\tau} \frac{[1-\cos(\omega\tau)]/|\tau|^{1+s} }{\sqrt{\int_{\omega'} G(\omega') [1-\cos(\omega'\tau)] }}.
\end{align}
This equation cannot be solved exactly for $G(\omega)$. However, we can extract the low-frequency behaviour of $G(\omega)$ by plugging an ansatz of the form $G^{-1}(\omega)=G_0^{-1} |\omega|^{1+2\zeta}$ in Eq.~\eqref{eq:var_opti} and matching the dominant low-frequency behaviours of both sides of the equation. To this end, we must first evaluate the integral in the denominator of Eq.~\eqref{eq:var_opti} which requires distinguishing a few cases.

If $\zeta>0$, one finds directly that
\begin{align}
    \int_\omega G_0\frac{1-\cos(\omega\tau)}{|\omega|^{1+2\zeta}}=-\frac{G_0}{\pi}|\tau|^{2\zeta} \Gamma(- 2\zeta) \cos(\zeta \pi).
\end{align}
Plugging this back into the Eq.~\eqref{eq:var_opti} creates a term
\begin{align}
     \frac{2\alpha}{\sqrt{- G_0 \Gamma(- 2\zeta) \cos(\zeta \pi)}}\int_{\tau_c}^\infty \d \tau \frac{1-\cos(\omega\tau)}{\tau^{1+s+\zeta}} \sim
 \begin{cases}
      |\omega|^{\zeta + s} \text{ if } \zeta < 2-s,\\
      |\omega|^2 {\tau_c}^{2-\zeta - s} \text{ if } \zeta> 2-s.
 \end{cases}
\end{align}
Matching the frequencies in Eq.~\eqref{eq:var_opti} therefore yields
\begin{itemize}
    \item $\zeta=s-1$ if $0<\zeta<2-s$, i.e. $1<s<3/2$,
    \item $\zeta=1/2$ if $\zeta>2-s$ and $\zeta>0$, i.e $s>3/2$.
\end{itemize}

If $\zeta<0$, one must reintroduce the UV cutoff $\tau_c$ to regulate the integral as
\begin{align}
    \underset{\omega\in[-\tau_c^{-1},\tau_c^{-1}]}{\int_\omega} G_0  \frac{1-\cos(\omega\tau)}{|\omega|^{1+2\zeta}} = \frac{G_0 \tau_c^{2\zeta}}{\pi} \int_0^1 \d y \frac{1-\cos(y \tau/\tau_c)}{y^{1+2\zeta}} \overset{\tau \gg \tau_c}{\longrightarrow} - \frac{G_0 \tau_c^{2\zeta}}{2 \pi \zeta}.
\end{align}
This creates the term
\begin{align}
    2\alpha\sqrt{-\frac{2\zeta}{G_0}}\int_{\tau_c}^\infty \frac{\d \tau}{\tau^{1+s}} \frac{1-\cos(\omega\tau)}{\tau_c^{\zeta}} \sim |\omega|^s. 
\end{align}
Going back to Eq.~\eqref{eq:var_opti}, one identifies
\begin{itemize}
    \item $\zeta=(s-1)/2$ if $\zeta<0$, i.e. $s<1$.
\end{itemize}
Gathering all cases yields Eq.~(13) in the main text.

\section{Numerical study of the LL to DP transition}
\label{SM:LL_DP_BKT}
\begin{wrapfigure}{r}{0.5\textwidth}
    \centering
    \vspace{-1cm}
    \includegraphics[width=\linewidth]{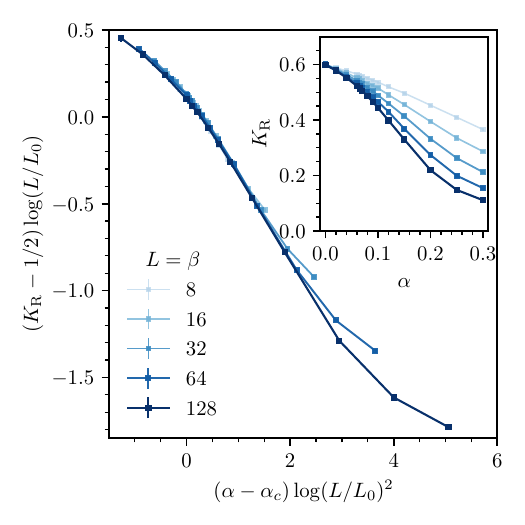}
    \caption{BKT finite-size scaling collapse for $K_{\rm R}=\pi \sqrt{\rho_s \kappa}$ at the DP to LL transition with $s=1$. Only $\alpha$ is varied while $K,u,g,s,\mu=0.6,1,0,1,1$. The optimal fitting parameters $L_0=1.33(3)$ and $\alpha_c=0.0605(5)$ were determined through the deviation from a local linear fit \cite{Kawashima_1993}. Inset: non-rescaled data with the LL at small $\alpha$ and DP at large $\alpha$. The convergence with system size to the universal jump at $\alpha_c$ is very slow as expected from the logarithmic scaling of Eq.~\eqref{eq:BKT_fss}.}
    \vspace{-1cm}
    \label{fig:BKT_KR}
\end{wrapfigure}
In this section, we consider the LL to DP transition which occurs for $s<3/2$. We decide to cross it by varying $\alpha$ and keeping all other parameters fixed. From the perturbative analysis presented in Sec.~\ref{SM:pert_analysis} we expect it to belong to the BKT universality class with a critical Luttinger parameter $K_c^{\rm LL-DP}=1-s/2$. However, it has also been suggested through a quantum Monte Carlo study conducted for $s=1$ in Ref.~\cite{Cai2014} that this transition is actually a second-order phase transition with dynamical critical exponent $z=2$. In the following, we reject the second-order scenario and validate the BKT one through precise finite-size scaling.

Following the standard BKT phenomenology \cite{Minnhagen_1987}, the renormalised Luttinger parameter $K_\R=\pi \sqrt{\kappa \rho_s}$ should, at the transition, exhibit a universal jump from $K_c^{\rm LL-DP}=1-s/2$ on the LL side to $0$ on the DP side. For a finite-size system with $L=\beta$, the jump is rounded according to the finite size scaling ansatz
\begin{equation}\label{eq:BKT_fss}
    K_\R(L,\alpha)-K_c^{\rm LL-DP}=l^{-1}F\left((\alpha-\alpha_c)l^2\right),
\end{equation}
where $\alpha_c$ is the critical point and $l=\log(L/L_0)$ with $L_0$ a non-universal length \cite{Harada_1997}. This suggests plotting the rescaled jump $l(K_\R-K_c^{\rm LL-DP})$ as a function of $(\alpha-\alpha_c)l^2$ with $\alpha_c$ and $L_0$ two fitting parameters. The resulting collapse for $s=1$ is shown in Fig.~\ref{fig:BKT_KR} and holds with great precision. This confirms that the DP to LL transition is indeed in the BKT universality class.

\vspace{1cm}

\section{Monte Carlo algorithm for the $(1+1)$D model}
\label{SM:MC_phi}

This section presents the Monte Carlo algorithm used to simulate the $(1+1)$D theory introduced in Eq.~\eqref{eq:phi_action}. We use a generalisation of the SmoWo (smooth worm) algorithm introduced in Ref.~\cite{bouvdup_2025} for the dissipationless case $\alpha=0$. The key idea of the SmoWo algorithm is to first map the model onto its  \emph{current-fluctuation} representation, i.e. a 2D model of loops and small fluctuations. The advantage of this representation is that loops can then be efficiently sampled using the worm algorithm \cite{Prokofiev_1998_worm,Prokofiev_2001_worm}, while the fluctuations are handled using the event-chain Monte Carlo (ECMC) \cite{Michel2014GenECMC,krauth2021ECMC}. In the following, we first derive the current fluctuation representation of the dissipative model \eqref{eq:phi_action}. We then give a brief description of the algorithm and refer the reader to Ref.~\cite{bouvdup_2025} for a more thorough discussion of the SmoWo algorithm and its performance.

\subsection{Current-fluctuation representation}
The model is put on a 2D lattice of size $L \times \beta \in \mathbb{N}^2$. The time and space lattice spacings play the role of the cutoffs $a$ and $\tau_c=a/u$. We work in dimensionless units where both spacings are set to $1$ and $u=1$. The field $\phi(x,\tau)$ is replaced by $\phi_i$ with $i\in \llbracket1,L\rrbracket \times \llbracket 1,\beta \rrbracket$ the site index. Discretising Eq.~\eqref{eq:phi_action} yields
\begin{align}\label{eq:action_phi_discr}
    S(\phi)=&\sum_i \Bigg\{\frac{1}{2\pi K} \left[\left(\phi_i-\phi_{i+\hat{x}} \right)^2 + \left(\phi_i - \phi_{i+\hat{\tau}} \right)^2\right] - g \cos(4\phi_i) + \frac{\mu}{\pi}  (\phi_{i+\hat{x}}-\phi_i) - \alpha \sum_{k=1}^{\beta-1} \cos(2\phi_i)\cos(2\phi_{i+k\hat{\tau}}) \mathcal{D}(k)\bigg\},
\end{align}
where $\hat{x}$ and $\hat{\tau}$ are unit vectors in the space and time directions. The long range kernel $1/|\tau|^{1+s}$ of the continuous model has been replaced by the $\beta$-periodic kernel
\begin{align}\label{eq:kernel}
    \mathcal{D}(j)=\frac{\sum_{n=1}^\beta e^{i \omega_n j}(1-\cos \omega_n)^{s/2}}{\sum_{n=1}^\beta e^{i\omega_n}(1-\cos \omega_n)^{s/2}},
\end{align}
with $\omega_n=2\pi n/\beta$. It is normalised such that $\mathcal{D}(1)=\mathcal{D}(\beta-1)=1$, decays as $\mathcal{D}(j)\sim 1/|j|^{1+s}$ for $1\ll j \ll \beta$. The reason this kernel is preferred over simpler alternatives such as $\mathcal{D}'(j)=\max(1/|j|^{1+s},1/|\beta-j|^{1+s})$ is that the Fourier transform of $\mathcal{D}(j)$ is $|\omega_n|^s$ for $\omega_n \ll 1$ down to the smallest frequency $\omega_1=2\pi/\beta$, while that of $\mathcal{D}'(j)$ only reproduces $|\omega_n|^s$ for $\omega_1 \ll \omega_n \ll 1$. The kernel \eqref{eq:kernel} thus minimises spurious finite-size effects appearing at low-frequency.

The boundary conditions for the discretised model are $\phi_i=\phi_{i+\beta\hat{\tau}} + \pi N_\tau$ and $\phi_i=\phi_{i+L\hat{x}} + \pi N_x$ with $N_\tau$, $N_x$ integers. This implies periodic boundary conditions for the fermionic field $\hat{\psi}$ appearing before bosonization. The partition function we wish to sample with our Monte Carlo algorithm is thus
\begin{align}
    Z=\sum_{N_x,N_\tau=-\infty}^{+\infty} \underset{\substack{\phi_i=\phi_{i+\beta\hat{\tau}} + \pi N_\tau\\ \phi_i=\phi_{i+L\hat{x}} + \pi N_x}}{\prod_i \int \d \phi_i} e^{-S(\phi)},
\end{align}
where we account for all winding numbers $N_x$ and $N_\tau$. The current-fluctuation representation of this model is obtained by writing first
\begin{align}
    \phi_i = \pi (n_i + f_i),
\end{align}
with the integer field $n_i\in \mathbb{N}$ and the fluctuations $f_i \in ]-1/2,1/2[$ such that Eq.~\eqref{eq:action_phi_discr} becomes
\begin{align}
    S(n,f)=&\sum_i \Bigg\{\frac{\pi}{2 K} \Big[(n_i - n_{i+\hat{x}}+f_i - f_{i+\hat{x}})^2 + (n_i - n_{i+\hat{\tau}} + f_i - f_{i+\hat{\tau}} )^2 \Big] \nonumber\\
    & - g \cos(4\pi f_i) + \mu (n_{i+\hat{x}} - n_i) - \alpha \sum_{k=1}^{\beta-1} \cos(2\pi f_i)\cos(2 \pi f_{i+k\hat{\tau}}) \mathcal{D}(k) \bigg\},
\end{align}
Next, we define the space-time current $\vec{J}_p=(J^x_p,J^\tau_p)$
\begin{align}\label{eq:current_def}
    J^x_p = n_{i(p)+\hat{\tau}} - n_{i(p)},\quad J^\tau_p = n_{i(p)}-n_{i(p)+\hat{x}},
\end{align}
where $i(p)=p-(\hat{x}+\hat{\tau})/2$. Physically, $\vec{J}$ captures the kinks of amplitude $\pi$ in the field $\phi$ and is thus the (coarse-grained) particle current. Since this current appears between lattice sites, e.g. $J^x_p$ is between the sites $i(p)+\hat{\tau}$ and $i(p)$, it connects the plaquettes $p$ of the lattice (hence the subscript in $\vec{J}_p$). It is furthermore conserved as $(\vec{\nabla} \cdot \vec{J})_p = J^x_{p+\hat{x}} - J^x_p + J^\tau_{p+\hat{\tau}} - J^\tau_p=0$ from the definitions above. In terms of this current, the discretised action finally takes its current fluctuation representation
\begin{align}\label{eq:current_fluct_action}
    S(\vec{J},f)= & \sum_p \Bigg\{ \frac{\pi}{2K} \Big[(J^\tau_p+ f_{i(p)} - f_{i(p)+\hat{x}} )^2 + (J^x_p + f_{i(p)+\hat{\tau}} - f_{i(p)})^2 \Big]\nonumber\\
    & - g \cos(4 \pi f_{i(p)} ) - \mu J^\tau_p - \alpha \sum_{k=1}^{\beta-1} \cos(2\pi f_{i(p)})\cos(2 \pi f_{i(p)+k\hat{\tau}}) \mathcal{D}(k) \bigg\}.\nonumber
\end{align}
Denoting by $\mathcal{C}_0$ the set of divergenceless current fields (i.e. the set of oriented closed loops), the partition function is
\begin{align}
    Z=\sum_{\vec{J}\in \mathcal{C}_0} \prod_i \int_{-\frac12}^{\frac12} \d f_i \, e^{-S(\vec{J},f)}.
\end{align}

\subsection{Algorithm}
The previous current-fluctuation representation separates loop degrees of freedom $\vec{J}$ from residual fluctuations $f$ which are respectively sampled with worm updates and ECMC updates. The worm moves require extending the configuration space by allowing for one open path, the worm, ranging from its tail at $p_t$ to its head at $p_h$. More formally, we call $\mathcal{C}_2(p_h,p_t)$ the set of divergenceless current fields except at $p_t$ where $(\vec{\nabla} \cdot \vec{J})_{p_t}=1$, and at $p_h$ where $(\vec{\nabla} \cdot \vec{J})_{p_h}=-1$. The extended partition function reads
\begin{align}
    Z_{\rm w}=\sum_{p_t,p_h} \sum_{\vec{J}\in \mathcal{C}_2(p_t,p_h)} \prod_i \int_{-\frac12}^{\frac12} \d f_i \, e^{-S(\vec{J},f)}.
\end{align}
A worm update then consists in either shifting the head $p_h \to p_h' \in \{p_h \pm \hat{x},p_h \pm \hat{\tau}\}$, or moving both the head and tail $p_h=p_t \to p_h'=p_t'$ if they are at the same position. All moves are accepted with the Metropolis filter. The physical configurations we sample are those with $p_h=p_t$ since the worm then disappears, and the intermediate configurations with a worm serve as paths between physical configurations. 

The fluctuations $f$ are updated with the ECMC algorithm. This requires further extending the configuration from $(\vec{J},f,p_h,p_t)$ to $(\vec{J},f,p_h,p_t,(e,i),(\varepsilon,\gamma),\sigma)$ where $e,\varepsilon \in \{-1,1\}$ are directions, $i,\in \llbracket 1, L \rrbracket \times \llbracket 1, \beta \rrbracket$ is a site label, $\gamma \in \partial p_h = \{ p_h \pm \hat{x}/2 \pm \hat{\tau}/2 \}$ is a site label close to $p_h$ and $\sigma \in \{0, 1\}$ denotes what type of update is being done. The ECMC also uses a continuous time $t$ instead of the discrete one usually found in Metropolis-like algorithms. When $\sigma=0$, the ECMC performs the continuous-time motion
\begin{equation}
    \partial_t f_j(t)=\begin{cases}
        e \text{ if } j=i,\\
        0 \text{ otherwise},
    \end{cases}
\end{equation}
while for $\sigma=1$,
\begin{equation}
    \partial_t f_j(t)=\begin{cases}
        \varepsilon \text{ if } j=\gamma,\\
        0 \text{ otherwise}.
    \end{cases}
\end{equation}
This deterministic evolution is interrupted by random events which update $(e,i)\to (e',i')$ in the first case, and $(\varepsilon,\gamma)\to (\varepsilon', \gamma')$ in the latter. The exact updates and the rates at which they occur are chosen such that the resulting algorithm is correct, i.e. satisfies the global balance condition. Although there are $O(\beta)$ rates, all rates can be examined in $O(\beta^0)=O(1)$ operations through the thinning procedure \cite{Lewis1979thinning,Michel_2019_clockMC}. We also add a worm event with the rate $\lambda_{\rm w}$ which changes $\sigma \to \sigma' = 1 - \sigma$ and attempts a worm move if $\sigma=0$. Finally, when $\sigma=0$ there is a refreshment event with rate $\lambda_{\rm r}$ and which uniformly resamples $(e,i)$. Typically, $\lambda_{\rm w}=1$ and $\lambda_{\rm r}= 0.1/(L\beta)$.

A pseudo-code implementation of the resulting SmoWo-LR algorithm for the model \eqref{eq:current_fluct_action} with long-range dissipative interactions is given in Alg.~\ref{alg:SmoWo_LR} where we used the notation $[x]_+=\max(0,x)$. It outputs $n_{\rm samples}$ samples separated by a time interval $T_{\rm samples}$. One can show using the infinitesimal generator formalism presented in the Appendix B of Ref.~\cite{bouvdup_2025} that it defines a valid (ergodic) Monte Carlo algorithm.

\begin{figure}[t]
\begin{minipage}[t]{0.49\columnwidth}
\begin{algorithm}[H]
\caption{SmoWo-LR}\label{alg:SmoWo_LR}
{\bf Input} $f, \vec{J}, i, e, \gamma, \varepsilon, \sigma, p_h, p_t, \lambda_{\rm r}, \lambda_{\rm w}, n_{\rm sample}, T_{\rm sample}$ \;
${\rm Sample} \gets \{\}$\;
$\bar \lambda = \{4\pi g, 4\pi \alpha \mathcal{D}(1), 4\pi \alpha \mathcal{D}(2), \dots, 4\pi \alpha \mathcal{D}(\beta-1)\}$\;
$\bar \lambda_{\rm tot}=\sum_{k=0}^{\beta-1}\bar \lambda[k]$\;
$t_s \gets T_{\rm sample}$\;
\While{$n_{\rm sample}>0$}{
    $i_\sigma = \sigma \gamma + (1-\sigma) i$\;
    $e_\sigma = \sigma \varepsilon + (1-\sigma) e$\;
    $t_{\rm r/w} \gets - \ln ({\rm ran}(0,1))/\lambda_{\rm r/w}$\; 
    $j_\q, t_\q \gets$ Alg.~\ref{alg:SR_quad}\;
    $j_\c, t_\c, s_\c \gets$ Alg.~\ref{alg:LR_cos}\;
    \uIf{$\sigma= 0$}{
        $t_{\rm event}={\rm min}(t_\q,t_\c,t_{\rm r},t_{\rm w})$\;
        ${\rm Sample} \gets$ Alg.~\ref{alg:sample_output}\;
        \uIf(\tcp*[f]{Quadratic event}){$t_\q =t_{\rm event}$}{
            $e,i \gets e,j_\q$\;}
        \uElseIf(\tcp*[f]{Cosine event}){$t_\c =t_{\rm event}$}{ 
            $e,i \gets - s_\c e, j_\c$\;}
        \uElseIf(\tcp*[f]{Refreshment}){$t_{\rm r} =t_{\rm event}$}{ 
            $i,e \gets {\rm choice}(\llbracket 1, L\rrbracket\times \llbracket 1, \beta \rrbracket \times \{-1,1\})$\;}
        \ElseIf(\tcp*[f]{Worm event}){$t_{\rm w} =t_{\rm event}$}{
            \uIf{$p_h=p_t$ and ${\rm ran}(0,1)<1/2$}{
                $p_h \gets {\rm Choice}(\llbracket 1, L\rrbracket\times \llbracket 1, \beta \rrbracket)$ \;
                $p_t \gets p_h$\;
                $\gamma, \varepsilon \gets {\rm choice}(\partial p_h \times \{-1,1\})$\;}
            \Else{ $p_h' \gets {\rm Choice}(\{p_h \pm \hat{x},p_h \pm \hat{\tau}\})$\;
                $P \gets \exp[S(p_h)-S(p_h')]$\;
                $P \gets  2^{\delta_{p_h,p_t}-\delta_{p_h',p_t}} P$\;
                \If{ ${\rm ran}(0,1)< P$}{
                $p_h \gets p_h'$\;
                $\gamma, \varepsilon  \gets {\rm choice}(\partial p_h \times \{-1,1\})$\;}}
            $\sigma \gets 1$\;}
        }
    \Else{
        $t_{\rm event}={\rm min}(t_\q,t_\c,t_{\rm w})$\;
        ${\rm Sample} \gets$ Alg.~\ref{alg:sample_output}\;
        \uIf(\tcp*[f]{Quad.}){$t_\q =t_{\rm event}$ and $j_\q \in \partial p_h$}{
            $\varepsilon,\gamma \gets \varepsilon, j_\q$\;}
        \uElseIf{$t_\q =t_{\rm event}$ and $j_\q \notin \partial p_h$}{ 
            $\varepsilon,\gamma \gets -\varepsilon, \gamma$\;}
        \uElseIf(\tcp*[f]{Cos.}){$t_\c = t_{\rm event}$ and $j_\c \in \partial p_h$}{ 
            $\varepsilon,\gamma \gets -s_\c \varepsilon, j_\c$\;}
        \uElseIf{$t_\c = t_{\rm event}$ and $j_\c \notin \partial p_h$}{ 
            $\varepsilon,\gamma \gets -\varepsilon, \gamma$\;}
    \ElseIf(\tcp*[f]{Worm event}){$t_{\rm w} =t_{\rm event}$}{
        $\sigma \gets 0$\;}
    }}
{\bf Return} ${\rm Sample}$\;
\end{algorithm}
\end{minipage}
\hfill
\begin{minipage}[t]{0.49\columnwidth}
\begin{algorithm}[H]
\caption{Ballistic motion $+$ Sampling}\label{alg:sample_output}
\uIf(\tcp*[f]{Output before event}){$t_s<t_{\rm event}$}{
    $f_{i_\sigma} \gets f_{i_\sigma} + e_\sigma \,t_s$\;
    \If{$p_h = p_t$}{
        Rebuild $\phi$ from $f,\vec{J}$\;
        ${\rm Sample}\gets {\rm Sample}\cup \{\phi\}$\;
        $n_{\rm sample} \gets n_{\rm sample}-1$\;
        }
        $f_{i_\sigma} \gets f_{i_\sigma} + e_\sigma \,(t_{\rm event}-t_s)$\;
        $t_s\gets t_s - t_{\rm event} + T_{\rm sample}$ \tcp*{time till next sample output}
    }
    \Else{$t_s\gets t_s- t_{\rm event}$\;
        $f_{i_\sigma} \gets f_{i_\sigma} + e_\sigma \,t_{\rm event}$\;}
{\bf Return} ${\rm Sample}$\;
\end{algorithm}

\begin{algorithm}[H]
\caption{Short-range quadratic event time}\label{alg:SR_quad}
$t_\q \gets +\infty$\;
$j_\q \gets i_\sigma$\;
$js = \{ i_\sigma +\hat{x}, i_\sigma -\hat{x}, i_\sigma + \hat{\tau}, i_\sigma - \hat{\tau}\}$\;
$p=i_\sigma +(\hat{x}+\hat{\tau})/2$\;
$Js=\{J_p^\tau, J_{p-\hat{x}}^\tau , -J_p^x, -J_{p-\hat{\tau}}^x \}$\;
\For{$n= 0,1,2,3$}{
    $y_\q \gets e_\sigma\left(f_{i_\sigma} - f_{js[n]} + Js[n]\right)$\;
    $t_\q^n \gets -y_\q + \sqrt{[y_\q]_+^2- \frac{2 K}{\pi}\ln ({\rm ran}(0,1)))}$\;
    \If{$t_\q^n < t_\q$}{
        $t_\q \gets t_\q^n$\;
        $j_\q \gets js[n]$\;
    }
}
{\bf Return} $j_\q, t_\q$\;
\end{algorithm}

\begin{algorithm}[H]
\caption{Long-range cosine event time}\label{alg:LR_cos}
    $t_\c \gets 0$\;
    \While{True}{
        $t_\c  \gets t_\c  -\ln ({\rm ran}(0,1)) / \bar \lambda_{\rm tot}$\;
        Pick $k \in \llbracket 0, \beta - 1 \rrbracket$ according to $\bar \lambda[k]/\bar\lambda_{\rm tot}$ using a Walker table\;
        \uIf{$k=0$} {   
        $s_\c \gets 1$\;
        }
        \Else{
        $s_\c \gets {\rm choice}(-1,1)$\;
        }
        \If{${\rm ran}(0,1)<[e_\sigma \sin(2\pi (f_{i_\sigma} + s_\c f_{i_\sigma+ k \hat{\tau}}))]_+$ } {
            {\bf Break}\;
        }
    }
    $j_\c \gets i_\sigma + k \hat{\tau}$\;
{\bf Return} $j_\c, t_\c, s_\c$\;
\end{algorithm}
\end{minipage}
\end{figure}

\section{Monte Carlo algorithm for the single-worldline theory}
\label{SM:MC_X}
This section is dedicated to the creation of an efficient Monte Carlo algorithm for the single-worldline theory. We describe the algorithm in detail before benchmarking its performance.

\subsection{Algorithm}
The field theory is put on a lattice with unit spacing and total length $\beta \in \mathbb{N}$. The field $X(\tau)$ thus becomes $X_i$ with $i=1,\dots,\beta$ and the discretised action is
\begin{align}\label{eq:action_X_discr}
    S(X)=&\sum_{i=1}^\beta \frac{1}{2}(X_i-X_{i+1})^2 + \frac{\alpha}{2} \sum_{i, j=1 (i \ne j)}^\beta |X_i-X_j| \mathcal{D}(i-j),
\end{align}
where the kernel $\mathcal{D}(j)$ has been introduced in the previous section.

We use the event-chain Monte Carlo (ECMC) algorithm (see \cite{Michel2014GenECMC,krauth2021ECMC} for an introduction) to perform a Monte Carlo simulation of the discrete model \eqref{eq:action_X_discr}. We start by lifting (i.e. extending) the configuration space to sample from $\Omega=\{X\}$ to $\Omega\times\mathcal{V}$ with $\mathcal{V}=\llbracket 1, \beta \rrbracket$ all site labels. The ECMC starts from a configuration $(X,i)\in\Omega \times \mathcal{V}$ and continuously evolves it through time as $(X(t),i(t))$. This evolution starts with the deterministic motion
\begin{equation}\label{eq:deterministic_motion}
    \partial_t X_j(t)=\delta_{i(t),j}, \quad \partial_t i(t)=0,
\end{equation}
where only $X_{i(t)}$ is updated. At any time $t$, the previous motion can be stopped by a random event that updates $i(t)\to j$. The deterministic motion then resumes with the new label $j$. Calling $i$ the current site variable, a random update to $j$ can be triggered by several independent Poisson rates which are listed below.
\begin{itemize}
    \item Each long-range interaction
    \begin{equation}
        S^{i,j}_{\rm LR}(X)=\alpha|X_i(t)-X_j|\mathcal{D}(i-j),
    \end{equation}
    can trigger the update $i \to j$ with the rate $\lambda_{\rm LR}^{i,j}(X(t))=[\partial_{X_i(t)} S^{i,j}_{\rm LR}]_+$ where $[x]_+=\max(0,x)$.
    \item The two short-range interactions
    \begin{equation}
        S^{i,i\pm 1}_{\rm SR}(X)=\frac{1}{2}(X_i(t)-X_{i\pm 1})^2,
    \end{equation}
    can trigger the update $i \to i \pm 1$ with the rate $\lambda_{\rm SR}^{i,i\pm 1}(X(t))=[\partial_{X_i(t)} S^{i,i\pm 1}_{\rm SR}]_+$.
    \item A refreshment event occurs with a constant rate $\lambda_r$ and uniformly picks a new label $j\in \mathcal{V}$ to replace $i$.
\end{itemize}
The first two events ensure that the Monte Carlo dynamics fulfils the global balance condition. This can be shown using the infinitesimal generator of the Markov process as in the Appendices of Ref.~\cite{bouvdup_2025}. The last refreshment event guaranties the irreducibility of the process, that is to say that any state $(X,i)$ can be reached from any other state $(X',i')$.

\begin{algorithm}[h!]
\caption{ECMC-LR}\label{alg:ECMC_LR}
{\bf Input} $X, n_{\rm sample}, T_{\rm sample}$ \;
${\rm Samples} \gets \{\}$\;
$i\gets {\rm choice}[1,\cdots,\beta]$\;
$t_s \gets T_{\rm sample}$\;
\While{$n_{\rm sample}>0$}{
    $j_{\rm SR} \gets \underset{j=i\pm1}{\rm argmin}(t_{\rm SR}^{i,j} \gets$ Eq.~\ref{eq:SR_event})\;
    $t_{\rm SR} \gets \underset{j= i \pm 1}{\rm min}(t_{\rm SR}^{i,j} \gets$ Eq.~\ref{eq:SR_event})\;
    $t_r \gets $ Eq.~\ref{eq:refreshment_event}\; 
    $t_{\rm LR} \gets 0$\;
    \While{True}{
        $t_{\rm LR}  \gets t_{\rm LR}  -\ln ({\rm ran}(0,1)) / \bar \lambda_{\rm LR}$\;
        Pick $k \in \llbracket 1, \beta\rrbracket$ according to $\bar \lambda_{\rm LR}^k/\bar\lambda_{\rm LR}$ using a Walker table\;
        \If{$X_i + t_{\rm LR} > X_{i+k}$}{
            {\bf Break}\;
          }  }
    $t_{\rm event}={\rm min}(t_{\rm SR},t_{\rm LR},t_r)$\;
    \uIf(\tcp*[f]{Outputting a sample}){$t_s<t_{\rm event}$}{      
        $X_i \gets X_i + t_s$\;
        ${\rm Samples}\gets {\rm Samples}\cup \{X\}$\; 
        $X_i \gets X_i + t_{\rm event}-t_s$\;
        $t_s\gets t_s - t_{\rm event} + T_{\rm sample}$\;
        $n_{\rm sample} \gets n_{\rm sample}-1$\;
        }
    \uElse{$t_s\gets t_s- t_{\rm event}$\;
        $X_i \gets X_i + t_{\rm event}$\;}
    \uIf(\tcp*[f]{SR event}){$t_{\rm SR} = t_{\rm event}$}{
        $i \gets j_{\rm SR}$ \;}
    \uElseIf(\tcp*[f]{LR event}){$t_{\rm LR} = t_{\rm event}$}{
        $i \gets i + k$\;}
    \uElse(\tcp*[f]{Refreshment event}){
        $i\gets {\rm choice}[1,\cdots,\beta]$\;}
}
{\bf Return} ${\rm Samples}$
\end{algorithm}

To numerically simulate this algorithm from the configuration $(X,i)$, we start by finding the times of the first occurrence of each event. The shortest of them, which we denote by $t_{\rm event}$ sets the stopping time for the deterministic dynamics in Eq.~\eqref{eq:deterministic_motion} so we update
\begin{equation}
    X_i \to X_i +t_{\rm event},
\end{equation}
and then $i\to j$ with $j$ given by the type of event. We then find the new minimal event time starting from this new configuration, and repeat the procedure over and over again. To compute the event time $t_\lambda$ associated with a rate $\lambda(X(t))$ we use inversion sampling : we draw  $\nu = {\rm ran}(0,1)$, a uniform number between 0 and 1 (one for each event), and solve
\begin{equation}
    \nu=\exp\left[-\int_0^{t_\lambda} \lambda(X(t)) \d t\right].
\end{equation}
This yields for the refreshment ($r$) and short-range (SR) events
\begin{align}\label{eq:refreshment_event}
    t_r&=-\frac{\log(\nu)}{\lambda_r},\\
    \label{eq:SR_event}
    t_{\rm SR}^{i,i\pm 1}&=X_{i\pm 1} -X_i +\sqrt{[X_i -X_{i\pm 1}]_+^2  - 2\log(\nu)}.
\end{align}
One could proceed similarly to get event times for the $\beta-1$ long-range events, which would require $O(\beta)$ operations to compute all of them. However, we are not interested in computing all long range event times as we only need to find the smallest. This can be done in $O(1)$ operations through a procedure known as thinning \cite{Lewis1979thinning,Michel_2019_clockMC} which relies on two key ideas: using properties of Poisson processes and bounding the event rates. In our case, we use the bound
\begin{align}
    \lambda_{\rm LR}^{i,j}(X(t))=[\partial_{X_i(t)} S^{i,j}_{\rm LR}]_+ \le \alpha \mathcal{D}(i-j)=\bar \lambda_{\rm LR}^{i-j},
\end{align}
and introduce $\bar \lambda_{\rm LR}=\sum_{k=1}^\beta \bar \lambda^k_{\rm LR}$. We then proceed as follows to find the smallest long-range event time $t_{\rm LR}$ along with the triggering site $j_{\rm LR}$.
\begin{enumerate}
    \item Start with $t_{\rm LR}=0$.
    \item While True:
    \begin{itemize}
        \item Draw $\nu = {\rm ran}(0,1)$ and set $t_{\rm LR} \to t_{\rm LR} - \log \nu /\bar \lambda_{\rm LR}$.
        \item Pick a distance $k$ according to the probability distribution $p(k)=\bar \lambda^k_{\rm LR}/\bar \lambda_{\rm LR}$. This can be done in $O(1)$ operations using Walker's method of alias \cite{Walker1977,Marsaglia2004gen_ran_var}.
        \item Break out of the ``While" loop with probability $\lambda^{i,i+k}_{\rm LR}(X_j + t_{\rm LR}\delta_{i,j})/\bar \lambda_{\rm LR}^k=\Theta(X_i + t_{\rm LR} - X_{i+k})$, i.e. if $X_i + t_{\rm LR} > X_{i+k}$.
    \end{itemize}
    \item The smallest long-range event time is $t_{\rm LR}$ and the triggering site is $j_{\rm LR}=i + k$.
\end{enumerate}
More details about this procedure can be found in Refs.~\cite{Michel_2019_clockMC,bouvdup_2025_bosonised1d}. 

With this continuous-time algorithm, computing the average of an observable $\mathcal{O}(X)$ is done by selecting a sampling time $T_{\rm sample}$ and using
\begin{equation}
    \langle \mathcal{O}(X)\rangle = \lim_{n_{\rm sample}\to \infty} \sum_{n=1}^{n_{\rm sample}} \frac{\mathcal{O}(X(t=n T_{\rm sample}))}{n_{\rm sample}}.
\end{equation}
Considering only the configurations at the event times would give a biased sampling and incorrect results. Piecing everything together, we arrive at the pseudo-code implementation detailed in Alg.~\ref{alg:ECMC_LR}.

\subsection{Performance test}
\begin{figure}[h!]
    \centering
    \includegraphics[width=0.5\linewidth]{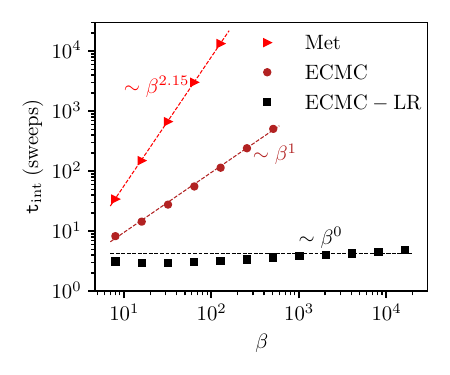}
    \caption{Integrated autocorrelation times $\texttt{t}_{\rm int}$ as a function of system size $\beta$ for $s=1$ and $\alpha=1$. The three scalings correspond to the Metropolis--Hastings algorithm (Met), the naive event-chain Monte Carlo algorithm (ECMC), and the ECMC adapted to long-range interactions (ECMC-LR).}
    \label{fig:t_int_X}
\end{figure}

Monte Carlo algorithms are efficient if, starting from a typical configuration $X$, they are able to quickly produce a new configuration $X'$ which is uncorrelated with $X$. On a more quantitative level, we consider the observable $\mathcal{O}(X)=|X(\omega_1=2\pi/\beta)|^2$ which we expect to decorrelate very slowly as it captures the large-scale physics. To define an objective measure of time to compare algorithms, we define the algorithmic time $\texttt{t}$ expressed in sweeps (i.e. $\beta$ operations) as increasing by $1/\beta$ each time an event time, event rate or pairwise interaction is computed. We then introduce the autocorrelation function
\begin{equation}
    C_\mathcal{O}(\texttt{t})=\frac{\langle {\cal O}(\texttt{t}){\cal O}(0)\rangle - \langle {\cal O}\rangle^2}{\langle {\cal O}^2\rangle - \langle {\cal O}\rangle^2},
\end{equation}
such that $C_\mathcal{O}(0)=1$ and $C_\mathcal{O}(+\infty)=0$, and which usually decays exponentially. Its characteristic timescale is extracted through the integrated autocorrelation time $\texttt{t}_{\rm int}=\frac12 \sum_\texttt{t} C_\mathcal{O}(\texttt{t})$. For a critical model, as is the case here, we expect $\texttt{t}_{\rm int} \sim \beta^{z_{\rm alg.}}$ with $z_{\rm alg.}\ge 0$ the algorithmic dynamical critical exponent which is non-zero in case of critical slowing down. We compare the ECMC-LR algorithm in Alg.~\ref{alg:ECMC_LR}, to an ECMC algorithm which naively (i.e. without thinning) computes all long-range event times, and to a Metropolis--Hastings (Met) algorithm which uniformly picks a site $i$ and updates $X_i \to X_i + \mathcal{N}(0,1)$ with the standard Metropolis rate. The results are shown in Fig.~\ref{fig:t_int_X} for $s=1$ and $\alpha=1$. The acceleration provided by both the ECMC and the thinning upgrade is clear. Their combination achieves $z_{\rm alg.}\simeq 0$, thus removing all critical slowing down. In more practical terms, the simulations performed in the main text for $\beta=2^{17}$ took a few days with the ECMC-LR algorithm. This very same task would have taken roughly $10^8$ years to complete with a simple Metropolis--Hastings algorithm.

\bibliography{refs}